\documentclass{acm_proc_article-sp}

\usepackage[utf8]{inputenc}
\usepackage[american]{babel}
\usepackage{graphicx}
\usepackage{color}
\usepackage{xspace}
\usepackage{amsmath}
\usepackage{ifthen}
\usepackage{subfigure}
\usepackage{morefloats}
\usepackage{balance}
\usepackage{amssymb}
\usepackage{numprint}
\usepackage{multirow}

\usepackage{footnote}
\usepackage[obeyspaces,hyphens]{url}
\usepackage{paralist}
\usepackage{soul}
\newcommand{\todo}[2][TODO]{\marginpar{\fbox{\parbox{.9\marginparwidth}{\raggedright\textbf{#1:} #2}}}}
\newcommand{\komm}[2]{\textbf{/* #2 (#1) */}}

\newcommand{\final}[1]{\textbf{/* for camera ready/long version: #1  */}}
\newcommand{\journal}[1]{\textbf{/* for journal version: #1  */}}

\newcommand{\story}[1]{\noindent\hrulefill\\\textbf{STORY:} \textit{#1}

\noindent\hrulefill}
\newcommand{\hypertext}[1]{#1}

\newcommand{\acknowledgment}[1]{#1}

 \renewcommand{\komm}[2]{} \renewcommand{\todo}[2][]{}
\renewcommand{\final}[1]{} 
\renewcommand{\journal}[1]{} 
\renewcommand{\story}[1]{}
\renewcommand{\acknowledgment}[1]{} % TODO: remove for final version

\newcommand{\smallparagraph}[1]{\smallskip\noindent\textbf{#1.}}
\newcommand{\assumptionsub}[1]{\smallskip\noindent\textbf{#1.}} % TODO we need to redefine this when we are closer to the submission. Use it for the different subsections of assumptions. Use the smallparagraph for the subsections of assumptionsubs

\newcommand{\assumptionintro}{\assumptionsub{Assumption}}
\newcommand{\assumptionevidence}{\assumptionsub{Evidence}}
\newcommand{\assumptionresults}{\assumptionsub{Results}}
\newcommand{\assumptiondiscussion}{\assumptionsub{Discussion}}

\newcommand{\captioncaption}[1]{\textbf{#1.}}

\newcommand{\eg}{e.g.,\xspace}
\newcommand{\etal}[1]{et~al.~\cite{#1}}
\newcommand{\ie}{i.e.,\xspace}
\newcommand{\Eg}{E.g.,\xspace}

\newcommand{\bibs}{BibSonomy\xspace}
\newcommand{\dogear}{Dogear\xspace}

\newcommand{\xmin}{\ensuremath{x_{min}}}

\makeatletter \@ifundefined{BibTeX}
   {\def\BibTeX{{\rmfamily B\kern-.05em%
    \textsc{i\kern-.025em b}\kern-.08em%
    T\kern-.1667em\lower.7ex\hbox{E}\kern-.125emX}}}{}
\makeatother

\newcommand{\furl}[1]{\footnote{\url{#1}}}
\newcommand{\egfurl}[1]{\footnote{e.g.~\url{#1}}}

\newcommand{\pears}{\ensuremath{r}\xspace}
\newcommand{\spear}{\ensuremath{\rho}\xspace}

\newcommand{\js}[1]{\ensuremath{JS_{#1}}}

\newcommand{\distSymbol}{D}
\newcommand{\distPEntityReqEntity}{\ensuremath{\distSymbol^{req}_{Entity}\xspace}}
\newcommand{\distPEntityReqFreq}{\ensuremath{\distSymbol^{req}_{F,Entity}\xspace}}
\newcommand{\distPEntityDBEntity}{\ensuremath{\distSymbol^{post}_{Entity}\xspace}}
\newcommand{\distPEntityDBFreq}{\ensuremath{\distSymbol^{post}_{F,Entity}\xspace}}

\newcommand{\distPTagReqEntityAll}{\ensuremath{\distSymbol^{req}_{Tag}\xspace}}
\newcommand{\distPTagReqFreqAll}{\ensuremath{\distSymbol^{req}_{F,Tag}\xspace}}
\newcommand{\distPTagReqEntityNonZero}{\ensuremath{^{\emptyset}\distSymbol^{req}_{Tag}\xspace}}

\newcommand{\distPTagDBEntityAll}{\ensuremath{\distSymbol^{post}_{Tag}\xspace}}
\newcommand{\distPTagDBFreqAll}{\ensuremath{\distSymbol^{post}_{F,Tag}\xspace}}
\newcommand{\distPTagDBEntityNonZero}{\ensuremath{^{\emptyset}\distSymbol^{post}_{Tag}\xspace}}

\newcommand{\distPUserReqEntityAll}{\ensuremath{\distSymbol^{req}_{User}\xspace}}
\newcommand{\distPUserReqFreqAll}{\ensuremath{\distSymbol^{req}_{F,User}\xspace}}
\newcommand{\distPUserReqEntityNonZero}{\ensuremath{^{\emptyset}\distSymbol^{req}_{User}\xspace}}

\newcommand{\distPUserDBEntityAll}{\ensuremath{\distSymbol^{post}_{User}\xspace}}
\newcommand{\distPUserDBFreqAll}{\ensuremath{\distSymbol^{post}_{F,User}\xspace}}
\newcommand{\distPUserDBEntityNonZero}{\ensuremath{^{\emptyset}\distSymbol^{post}_{User}\xspace}}

\newcommand{\distPPubReqEntityAll}{\ensuremath{\distSymbol^{req}_{Pub}\xspace}}
\newcommand{\distPPubReqFreqAll}{\ensuremath{\distSymbol^{req}_{F,Pub}\xspace}}
\newcommand{\distPPubReqEntityNonZero}{\ensuremath{^{\emptyset}\distSymbol^{req}_{Pub}\xspace}}

\newcommand{\distPPubDBEntityAll}{\ensuremath{\distSymbol^{post}_{Pub}\xspace}}
\newcommand{\distPPubDBFreqAll}{\ensuremath{\distSymbol^{post}_{F,Pub}\xspace}}
\newcommand{\distPPubDBEntityNonZero}{\ensuremath{^{\emptyset}\distSymbol^{post}_{Pub}\xspace}}

\newcommand{\threefig}[8] {
	\begin{figure*}
		\centering
		\subfigure[#2.]{
			 \centering
			 \includegraphics[width=0.3\textwidth]{figs/#1}
			 \label{fig:#1}
                       }
		\subfigure[#4.]{
			 \centering
			 \includegraphics[width=0.3\textwidth]{figs/#3}
			 \label{fig:#3}
                       }
        \subfigure[#6.]{
			 \centering
			 \includegraphics[width=0.3\textwidth]{figs/#5}
			 \label{fig:#5}
                       }
		\caption{#8.}
		\label{fig:#7}
	\end{figure*}
}

\newcommand{\logLoggedInNoSpamDBNoSpamTagEntityAllPearsonCorrelation}{\numprint{0.4200243320161999}}

\newcommand{\logLoggedInNoSpamDBNoSpamTagEntityAllSpearmanCorrelation}{\numprint{0.05853252770732}}

\newcommand{\logLoggedInNoSpamDBNoSpamTagEntityAllJS}{\numprint{0.4409793693099853}} %%%% These macros have been created using class org.bibsonomy.analysis.tag.EntityFrequencyCounting and they INCLUDE references to undefined categories. These however occur only rarely for logged-in non-spammers. Please confer the log files to check on that. 

\newcommand{\logLoggedInNoSpamDBNoSpamTagEntityNonZeroPearsonCorrelation}{\numprint{0.4145511207238073}}

\newcommand{\logLoggedInNoSpamDBNoSpamTagEntityNonZeroSpearmanCorrelation}{\numprint{0.5152841499219215}}

\newcommand{\logLoggedInNoSpamDBNoSpamTagEntityNonZeroJS}{\numprint{0.2716763787626671}}
\newcommand{\logLoggedInNoSpamDBNoSpamTagFreqAllPearsonCorrelation}{\numprint{0.9684883835698082}}

\newcommand{\logLoggedInNoSpamDBNoSpamTagFreqAllSpearmanCorrelation}{\numprint{0.62030572218825}}

\newcommand{\logLoggedInNoSpamDBNoSpamTagFreqAllJS}{\numprint{0.050929553428304686}} %%%% These macros have been created using class org.bibsonomy.analysis.tag.EntityFrequencyCounting and they INCLUDE references to undefined categories. These however occur only rarely for logged-in non-spammers. Please confer the log files to check on that. 

\newcommand{\logLoggedInNoSpamDBNoSpamUserEntityAllPearsonCorrelation}{\numprint{0.09249468279998238}}

\newcommand{\logLoggedInNoSpamDBNoSpamUserEntityAllSpearmanCorrelation}{\numprint{0.5722747090523841}}

\newcommand{\logLoggedInNoSpamDBNoSpamUserEntityAllJS}{\numprint{0.49024896815887026}} %%%% These macros have been created using class org.bibsonomy.analysis.tag.EntityFrequencyCounting and they INCLUDE references to undefined categories. These however occur only rarely for logged-in non-spammers. Please confer the log files to check on that. 

\newcommand{\logLoggedInNoSpamDBNoSpamUserEntityNonZeroPearsonCorrelation}{\numprint{0.08092343672337272}}

\newcommand{\logLoggedInNoSpamDBNoSpamUserEntityNonZeroSpearmanCorrelation}{\numprint{0.7177672507146746}}

\newcommand{\logLoggedInNoSpamDBNoSpamUserEntityNonZeroJS}{\numprint{0.4709815103621574}}
\newcommand{\logLoggedInNoSpamDBNoSpamUserFreqAllPearsonCorrelation}{\numprint{0.9416442341710409}}

\newcommand{\logLoggedInNoSpamDBNoSpamUserFreqAllSpearmanCorrelation}{\numprint{0.21018762832811064}}

\newcommand{\logLoggedInNoSpamDBNoSpamUserFreqAllJS}{\numprint{0.19448221229265805}} %%%% These macros have been created using class org.bibsonomy.analysis.tag.EntityFrequencyCounting and they INCLUDE references to undefined categories. These however occur only rarely for logged-in non-spammers. Please confer the log files to check on that. 

\newcommand{\logLoggedInNoSpamDBNoSpamResBibtexEntityAllPearsonCorrelation}{\numprint{0.5525678179384187}}

\newcommand{\logLoggedInNoSpamDBNoSpamResBibtexEntityAllSpearmanCorrelation}{\numprint{0.03031791242265086}}

\newcommand{\logLoggedInNoSpamDBNoSpamResBibtexEntityAllJS}{\numprint{0.7066272363568242}} %%%% These macros have been created using class org.bibsonomy.analysis.tag.EntityFrequencyCounting and they INCLUDE references to undefined categories. These however occur only rarely for logged-in non-spammers. Please confer the log files to check on that. 

\newcommand{\logLoggedInNoSpamDBNoSpamResBibtexEntityNonZeroPearsonCorrelation}{\numprint{0.6084357144320965}}

\newcommand{\logLoggedInNoSpamDBNoSpamResBibtexEntityNonZeroSpearmanCorrelation}{\numprint{0.24780783678354795}}

\newcommand{\logLoggedInNoSpamDBNoSpamResBibtexEntityNonZeroJS}{\numprint{0.15124206606795065}}
\newcommand{\logLoggedInNoSpamDBNoSpamResBibtexFreqAllPearsonCorrelation}{\numprint{0.8041792208572317}}

\newcommand{\logLoggedInNoSpamDBNoSpamResBibtexFreqAllSpearmanCorrelation}{\numprint{0.7783262880575688}}

\newcommand{\logLoggedInNoSpamDBNoSpamResBibtexFreqAllJS}{\numprint{0.34419570520460274}} %%%% These macros have been created using class org.bibsonomy.analysis.tag.EntityFrequencyCounting and they INCLUDE references to undefined categories. These however occur only rarely for logged-in non-spammers. Please confer the log files to check on that. 

\newcommand{\totalNumUsers}{17932}
\newcommand{\totalNumBookmarkPosts}{551606}
\newcommand{\totalNumPublicationPosts}{2391721}
\newcommand{\totalNumTags}{250344}
\begin{document}

\newcommand{\papertitle}{Of course we share! Testing Assumptions about Social Tagging Systems}
\newcommand{\papersubtitle}{A Study of User Behavior in BibSonomy}

\pdfinfo{
/Title (\papertitle - \papersubtitle)
/Author (Stephan Doerfel, Daniel Zoller, Philipp Singer, Thomas Niebler, Andreas Hotho, Markus Strohmaier)
/Keywords(social tagging, assumptions, social sharing, folksonomy, bookmarking,
 tagging, behavior)
}

\title {\papertitle}
\subtitle{\papersubtitle}

\numberofauthors{6} 
\author{\alignauthor
Stephan Doerfel\\
       \affaddr{University of Kassel}\\
       \affaddr{Kassel, Germany}\\
       \email{doerfel@cs.uni-kassel.de}
\alignauthor
Daniel Zoller\\
       \affaddr{University of Würzburg}\\
       \affaddr{Würzburg, Germany}\\
       \email{zoller@informatik.uni-wuerzburg.de}
\alignauthor
Philipp Singer\\
       \affaddr{Technical University Graz}\\
       \affaddr{Graz, Austria}\\
       \email{philipp.singer@tugraz.at}
\and
\alignauthor
Thomas Niebler\\
       \affaddr{University of Würzburg}\\
       \affaddr{Würzburg, Germany}\\
       \email{niebler@informatik.uni-wuerzburg.de}
\alignauthor
Andreas Hotho\\
       \affaddr{University of Würzburg}\\
        \affaddr{Würzburg, Germany}\\
       \email{hotho@informatik.uni-wuerzburg.de}
\alignauthor
Markus Strohmaier\\
       \affaddr{GESIS \& U. of Koblenz}\\
       \affaddr{Cologne, Germany}\\
       \email{markus.strohmaier@gesis.org}
}

\maketitle

\begin{abstract}

Social tagging systems have established themselves as an important part in
today's web and have attracted the interest of our research community in a
variety of investigations.
Henceforth, several assumptions about social tagging systems have
emerged on which our community also builds their work. Yet, testing such assumptions has
been difficult due to the absence of suitable usage data in the past. In this
work, we thoroughly investigate and evaluate four assumptions about tagging systems -- covering social,
retrieval, equality, and popularity issues --
by examining live server log data gathered from the real-world, public social tagging system \bibs. Our empirical results indicate that while some
of these assumptions hold to a certain extent, other assumptions need
to be reflected in a very critical light. Our observations have
implications for the understanding of social tagging systems, and the way they are used in open environments.

\end{abstract}



%%% Local Variables:
%%% mode: latex
%%% TeX-master: "paper"
%%% End:

 \maketitle

\nprounddigits{3}
\npdecimalsign{.}
\npfourdigitnosep
\section{Introduction}
\label{sec:intro}

Social tagging systems such as Delicious, \bibs or Flickr have
attracted the interest of our research community for almost a
decade~\cite{Mathes2004,golder2006usage}. Significant advances have
been made with regard to our understanding about the emergent,
individual and collective processes that can be observed in such
systems~\cite{strohmaier2010users}.  Useful algorithms for
retrieval~\cite{hotho2006information} and
classification~\cite{zubiaga2011shelves} have been developed that
exploit the rich fabric of \hypertext{links between users, resources,
  and tags in} social tagging systems for facilitating information
organization, search and navigation. Other work has focused on the
extraction or stabilization of emergent
semantics~\cite{golder2006usage,cattuto2007network}.

While this line of research has significantly increased our ability to
describe, model, and utilize social tagging systems, our community has
also built their work on certain assumptions about how these systems
are used, which have emerged over time. For such assumptions, arguments and
evidence have been discussed in literature. Yet, due to a lack of appropriate
data and other issues, these assumptions have gone largely unchallenged and it
is unclear to which degree they do hold in actual tagging systems. Some of
these assumptions are controversial and researchers have argued for and
against them in our community, providing thus all the more reason to evaluate
them on real-world usage data. Only a few studies have analyzed user behavior
in social tagging systems to better understand such assumptions, either by
(i)~conducting user surveys (\eg by Heckner~\etal{heckner2009personal}) or by
(ii)~tapping into the rich corpus of tagging data that is available on the web
(\ie the posts) (\eg by Cattuto~\etal{cattuto2007network}). However, such
studies come with certain limitations such as self-reporting biases or the
lack of detailed usage data -- \ie how users actually request
information. In this paper we overcome these drawbacks by presenting
and thoroughly investigating a detailed usage log of a popular social
tagging system. This allows us to test and challenge a series of
assumptions from related work leveraging usage data of the real-world,
open social tagging system \bibs.\furl{http://www.bibsonomy.org/}

\smallparagraph{Research questions} 
We discuss and evaluate the following four controversial assumptions about tagging:
(i)~The \emph{social assumption} establishes that tagging systems are supposed
to be used collaboratively to tag and share resources. We investigate to which
degree such sharing actually happens and discuss evidence for the interest of
users in the content of others.
(ii)~The \emph{retrieval assumption} states that users tag resources for later
retrieval.
(iii)~The \emph{equality assumption} claims, that the three sets of
entities in a tagging system -- users, tags and resources -- are equally
important. This assumption is inherent in the folksonomy model
(\eg~\cite{hotho2006information}) that is a popular basis for recommendations
and ranking algorithms. 
(iv)~The \emph{popularity assumption} suggests that the popularity of
users, tags, and resources in posts is matched by their popularity in
retrieval. In tagging systems, popularity is used for example in tag
clouds where frequent tags have large font sizes to gain the users'
attention and to be easily accessible by a mouse click.

\smallparagraph{Findings} In our analysis of the social tagging system
\bibs, we find evidence both for and against the \emph{social assumption}.
While some user actions indeed indicate social sharing purposes, others are
evidence for individual purposes, suggesting that social tagging systems
provide utility because they can support both kinds of modes in a flexible
manner. We also find that while users post a large number of resources and tags
to \bibs, they only retrieve a rather small fraction of them later, which
provides first evidence that the \emph{retrieval assumption} might not hold for
systems such as \bibs. Next, we find a strong inequality between the use of
users, tags and resources within \bibs. User pages are visited much more
often than corresponding resource or tag pages, providing clear evidence that
the \emph{equality assumption} in \bibs is wrong. Finally, while we observe
common usage patterns in post and request behavior on an aggregate level, the
patterns are less pronounced on an individual level, suggesting that the
\emph{popularity assumption} only holds to certain extents.

\smallparagraph{Contributions} The paper makes contributions on three
levels. \emph{(i)~Methodical:} We identify a series of assumptions and
illuminate a way towards testing them with log data. While our
findings are limited to a single system (\bibs), our method of testing
these assumptions is general. The approach can well be applied to other
social tagging systems to test the extent to which these assumptions
hold in different contexts. \emph{(ii)~Empirical:} We challenge a
number of assumptions by testing them with actual log data and report
their validity for the popular social tagging system
\bibs. \emph{(iii)~Data driven:} We make an anonymized \bibs server
log dataset available to other researchers (see Section~\ref{sec:dataset}). This
will enable our community to investigate similar or different questions on a
unique dataset that has not been available yet.

\smallparagraph{Structure of this paper} After the discussion of
related work in Section~\ref{sec:relwork}, we describe the \bibs
datasets in Section~\ref{sec:dataset}. We then turn our attention on
studying and \hypertext{evaluating} the aforementioned four
assumptions on social tagging in Section~\ref{sec:assumptions}. \hypertext{In
Section~\ref{sec:discussion} we discuss differences between \bibs
and other tagging systems.} Finally, Section~\ref{sec:conclusions} concludes the paper.

Overall, our findings are relevant for researchers interested in user behavior
and modeling in the context of social tagging systems and their adoption as
well as to system engineers interested in improving the utility and usefulness
of social tagging systems on the web.

%%% Local Variables: 
%%% mode: latex
%%% TeX-master: "paper"
%%% End: 

\section{Related Work}
\label{sec:relwork}
In this section we discuss related literature on the investigation of
tagging systems and log file analysis in general. Further related
work, that is specifically relevant to individual assumptions, will be
discussed in greater detail later in the corresponding context
\hypertext{in Section~\ref{sec:assumptions}}.

\smallparagraph{Inception} Work on social tagging and emerging folksonomies
began in late 2004, when the term \emph{folksonomy} was first mentioned by
Vander Wal\furl{http://vanderwal.net/random/category.php?cat=153} and
continued in 2005 in various blog posts and papers.
One of the first reviews about social tagging systems was provided by
Mathes in~\cite{Mathes2004}. He noted that social tagging systems allow a much greater
variability in organizing content than formal classification can provide.
Mathes was also among the first to hypothesize that tag distributions
may emerge to power law distributions which can characterize the
semantic stabilization of such systems (see
also~\cite{golder2006usage}). Furthermore, Mathes identified some
potentials and uses of tagging systems, such as serendipitous
browsing.

\smallparagraph{User Surveys and Post Analysis} 
Abrams~\etal{abrams1998information} already discussed the management of website
bookmarking long before the rise of social tagging on the Web using a user
survey and bookmark files from participants. Their results showed that users are
motivated to share bookmarks (still via email back then) as well as to retrieve
them later.
Heckner \etal{heckner2009personal} conducted a survey of tagging
systems (namely Flickr, YouTube, Delicious and Connotea) with 142
users regarding their motivations. The results showed that there are
mainly two motivations for users to post content: sharing resources
with others and storing them primarily for personal retrieval later
on. The strength of these motivations varies from system to system.

Using the post data of tagging systems, several studies analyzed
aspects of posting behavior, \eg the distributions of users,
resources, and tags in posts \cite{cattuto2007network}, or the
identification of different types of users -- categorizers and
describers -- regarding their choice of tags
\cite{strohmaier2010users}.  However, these studies did not use log
data for their analysis to explore the \hypertext{actual} retrieval
behavior. A review of social tagging regarding a variety of
  diverse aspects of such systems -- including vocabulary, structure,
  visualization, motivation, or search and ranking -- was created by
  Trant~\cite{trant2009studying}.

\smallparagraph{Web Log Mining} Predominantly, web logs have been used
to investigate the query behavior in search engines or the usage of
digital libraries in order to better understand a system's users. This
can help webmasters to tailor their systems more specifically to the
users' needs. A survey on such works about search engines
is given by Agosti \etal{agosti2012analysis}. Examples
  for the analysis of digital libraries can be found in the works of
  Nicholas et al. (\eg \cite{nicholas2005scholarly}). Tagging systems
exhibit aspects of both search engines and libraries. While they are
collections of resources with description and categories,
however not professionally organized like in a digital library, they
are organized by users in their individual fashion of assigning tags
and entering meta data. Nonetheless, the data is clearly more
structured than data on the Web in general as posts are constructed
according to a specified template.

 Carman~\etal{carman2009statistical} combine tagging data with
 log data from search engines and compare the distribution of tags to
 that of query terms in search. They find a large overlap in the
 systems' vocabularies and correlations between the frequency
 distributions of queries and tags to the same URLs. However, they
 also provide evidence that both tag and query term samples do not
 come from the same distribution.

While there exists a large amount of literature on tagging
  systems, to the best of our knowledge, the only work utilizing and
  analyzing log data from a tagging system are those by Millen
  and Feinberg~\cite{Millen2006}, Millen~\etal{Millen2007social}, and by Damianos~\etal{Damianos2007}.
Millen and Feinberg investigated user logs of the social tagging
system \emph{\dogear{}} (internally used at IBM) with a focus on
social navigation in the system.
They found strong evidence that social navigation -- \ie users who are
regularly looking at bookmark collections of other people -- is a
fundamental part of the social tagging system. They also found a
positive correlation between the assignment frequency of a tag in
posts and the frequency of it being used for browsing.
These findings have been highly relevant for the
  understanding of tagging behavior as they provide actual evidence of
  how users make use of a tagging system's content.
  Millen~\etal{Millen2007social} combined log analysis and user
  interviews to investigate the way users retrieve resources. They
  observe diverse behavior patterns for different users and find
  that heavy users tend to spend more time with their own collections than
  user with only few bookmarks. Damianos~\etal{Damianos2007}
  introduced a tagging prototype called \emph{onomi} to the
  organization MITRE. They use log data to determine how well the
  system was accepted and present several usage statistics from a ten
  month test period. They found that their users can be categorized
  into information providers and information consumers depending on
  their individual ratio of browsing and bookmarking activities.

We compare findings in our experiments to the above
  mentioned analyses where possible. However, all three works focus on
  local social tagging systems located inside the network of a
  particular company. Therefore, they represent
private systems where users only tag resources inside the company's
field of interest and hence, the results are hard to compare with a
real world tagging system. Millen~\etal{Millen2007social} already note, that these kinds of
  services require their users to use corporate identities instead of
  pseudonyms, which is typically not the case in public systems.
Contrary, in this work we focus on the publicly available system \bibs
to overcome this limitation. This leads to some interesting deviating
insights that are discussed in Section \ref{ssec:as:tagcloud}
regarding the social and the popularity assumption. While we not only
extend the analyses in~\cite{Millen2006,Millen2007social,Damianos2007} by investigating a series of assumptions
about social tagging systems, we also benefit from long-time log data
allowing us to get a clearer overview over actual user behavior in an
already established social tagging system.

%%% Local Variables: 
%%% mode: latex
%%% TeX-master: "paper"
%%% End: 

\urldef{\helppage}\url{http://www.bibsonomy.org/help_en/URL Scheme Semantics}
\section{Dataset}

\label{sec:dataset}
The datasets used in this paper are based on web server logs and database
contents of the social tagging system \bibs~\cite{benz2010social}. \bibs allows
users to store, tag, and share links to websites and (scientific) publications
and offers for example the following options to query for posts:\footnote{For
details on the \bibs URL schema see \helppage.} A user can request to see all
posts with one or several tags, or posts from a specific user or group, or use
a combination of user and tag restrictions. For each resource, \bibs has a page
that lists its tags and users from all posts. Publication posts have a
\emph{details} page that shows the meta data of the publication (as entered by
the user who created the post) and offers export options. Posts of bookmarked
websites can also contain meta data (like a description of the website), but
requests to a bookmark are usually conducted by just clicking on a post's title
to reach the website. Such requests are not recorded in \bibs's server logs and
therefore, we must restrict some experiments exclusively to publication requests.

In \bibs, users can form groups or declare friendship to other users. Both
friendships and groups are used in the visibility concept of posts. \bibs
offers many further features like discussion forums, or a full text search,
that exceed the usual tagging system functionality. Therefore, such features
have been excluded from our experiments.

Due to its high rank in search engines, \bibs is a popular target for spammers.
Spammers are users who store links to advertisement sites to increase their
visibility on the web.
\bibs uses a
learning classifier~\cite{krause2008antisocialb} as well as manual
classification by the system's administrators to detect spam. In all
experiments, we only used data of users that have been classified as
non-spammers.

We restricted the datasets to data that had been created between the start of
\bibs in 2006 and the end of 2011, since early in 2012 the login mechanism was
modified, which introduced significant changes to the logging infrastructure.
\emph{With this paper, we make anonymized datasets of logs and posts available
to researchers.}\furl{http://www.kde.cs.uni-kassel.de/bibsonomy/dumps/}

\smallparagraph{User and Content Dataset}
We use tagging data from \bibs's database, \ie the users with their posts,
containing resources and tags, as well as all data about groups and
friendships. In the considered time frame, \npnoround\numprint{852172} people
registered a user account of which \npnoround\numprint{\totalNumUsers} were
classified as non-spammers. They created
\npnoround\numprint{\totalNumBookmarkPosts} bookmark posts and
\npnoround\numprint{\totalNumPublicationPosts} publication posts using
\npnoround\numprint{\totalNumTags} tags.

\smallparagraph{Request Log Dataset}
The \bibs log files include all HTTP requests (caching is disabled) to the system 
including common request attributes like IP address, user agent, date, and
referer, as well as a session identifier and a cookie containing the name of
the logged-in user. Out of the over $2.5$ billion requests, we used only those
from logged-in non-spammers and additionally filtered out requests to extra
resources including CSS, JavaScript, and images files as well as requests
from web bots (comparing user agents to those of known bots in various online
databases). Finally, after removing requests to pages that are irrelevant to our
study (like help pages and administration pages), the remaining dataset
contained about $3.5$ million requests.
%%% Local Variables: 
%%% mode: latex
%%% TeX-master: "paper"
%%% End: 

\section{Assumptions}\label{sec:assumptions}
In this section, we present our results. For each assumption, we (i) make the assumption explicit, (ii) provide evidence for the assumption in the literature, (iii) present the results of our research and (iv) discuss our findings.

\clearpage
\subsection{The Social Assumption}
\label{ssec:socialness}

\assumptionintro{}
The \emph{social assumption} \hypertext{states} that users of
social tagging systems use the system to (re)use resources that have been shared and tagged
by others, either by viewing them or by copying them into
their own collection.

\assumptionevidence{} The social aspect of tagging has been
subject to controversial discussion in the past. It has
  been praised and disputed already early in the history
of tagging systems. Mathes~\cite{Mathes2004} stated that folksonomies
could ``lower the barriers to cooperation'' and
Weinberger~\cite{weinberger2005tagging} named it as one of two aspects
that ``make tagging highly useful''.  Marlow~\etal{Marlow2006}
presented an early model for social tagging systems where they argued
that social relations between users are a critical element. The
authors point out that social interaction connects bookmarking
activities of individuals with a rich network of shared tags,
resources and users.
Furthermore, Millen and Feinberg~\cite{Millen2006} found out that around $74\%$ of all page
requests in \dogear{} -- an internal social tagging service at IBM --
refer to content that was bookmarked by other users. In contrast
to that, Damianos~\etal{Damianos2007} noticed that users are looking more at
their own (70\%) than at other users' collection in their system onomi. Yet, it is not self-evident
that similar observations can be made for public tagging systems, where users use the system without direct company
guidance that might influence their behavior.
Contrary, users may choose to use such systems for individual purposes
only, \eg to create their own collections, and thus ignore the
resources of other users.  Vander Wal~\cite{VanderWal2005} already
pointed out that personal information management may be one of the
main reasons why people use social tagging systems which was also
emphasized by Terdiman~\cite{Terdiman2005}. Porter~\cite{Porter2005b}
claimed that ``Personal value precedes network value: Selfish use
comes before shared use''.
A user survey by Heckner~\etal{heckner2009personal} found that about
$70\%$ of all users store resources in tagging systems comparable to
\bibs mainly to retrieve them themselves, not particularly to share
them (in contrast to systems where images or videos are shared). 
However, it is also noted that ``even users of systems who claim that personal information
management is very important for them, state that
sharing is also part of their motivation of using the systems''~\cite{heckner2009personal}.
While this survey takes the perspective of \emph{motivation} for posting we
will rather take the viewpoint of the \emph{usage} of posts. We
conducted a first evaluation of social behavior in
\cite{doerfel2014social}, which we extend and detail in the following.

\assumptionresults{}
\emph{Visiting Content.} First, we analyze the ownership of visited (retrieved) content.
Table~\ref{tab:request_own_other}
shows the number of requests to pages for different ownership
categories.\footnote{\label{foot:landing}The \bibs landing page was considered
separately because, although it lists recent posts of any users, many users
just visit it to start retrieval by using the provided input fields on the
page, and thus ignore the displayed resources.}
We can observe that more than two thirds of all requests of logged-in
users target their own pages. Users visit other pages in about~$32\%$
of the requests to look at either general pages, \ie pages containing
posts of several users (about $16\%$), or content of individual other users
or groups (about $16\%$). Hereby, requests to groups and friend pages are
both rather infrequent (about~$3\%$) indicating that these
particularly social features (in \bibs they are used to control the
visibility of posts) play only a minor role. Further, the share of
visits to content of others is below the reported $74\%$ for a company
internal tagging system by Millen and Feinberg~\cite{Millen2006}, but
similar to the reported share by Damianos~\etal{Damianos2007}. In summary, we
see that the larger share of interactions in \bibs happens with the personal collection.  However, the interest in other users' content
accounts for a significant part -- almost one third of all retrieval
requests -- of the interaction with the system.
\begin{table}[b!]
\centering
\caption{\textbf{Content visits.} Request counts to the
(logged-in user's) own content, to content from other users, or to general (non-user-specific) pages. Requests
to the landing page (see Footnote~\ref{foot:landing}) are not considered in the percentage
calculation.
}
\begin{tabular}{c|r|r}
 & \# request & \% requests \\
\hline
user's own & \nprounddigits{0}\numprint{1018089.0} &
\nprounddigits{2}\numprint{68.36638762904875}\\
%\hline
groups and friends & \nprounddigits{0}\numprint{44875.0} &
\nprounddigits{2}\numprint{3.013431679208362}\\
%\hline
other users & \nprounddigits{0}\numprint{190394.0}
&\nprounddigits{2}\numprint{12.785277128271796}\\
%\hline
general & \nprounddigits{0}\numprint{235808.0} &
\nprounddigits{2}\numprint{15.834903563471098}\\
%\hline
landing page & \nprounddigits{0}\numprint{296788.0} &
-\\
\end{tabular}
\label{tab:request_own_other}
\end{table}

\newcommand{\couldBeCopied}{\nprounddigits{0}\numprint{16,697625}} % could be copied / all post actions
\newcommand{\indeedCopied}{\nprounddigits{0}\numprint{41,795152}}

\newcommand{\totalBookmarklet}{\nprounddigits{0}\numprint{64.11200}}
\newcommand{\totalManually}{\nprounddigits{0}\numprint{24.99600}}
\newcommand{\totalCopy}{\nprounddigits{0}\numprint{10.70600}}
\newcommand{\publCopy}{\nprounddigits{0}\numprint{17.61000}}
\newcommand{\bookCopy}{\nprounddigits{0}\numprint{3.46400}}
\newcommand{\bookBookmarklet}{\nprounddigits{0}\numprint{80.72400}}
\newcommand{\publBookmarklet}{\nprounddigits{0}\numprint{48.27500}}

\emph{Copying Resources.}  When users added new posts to their
collections, in \totalCopy\% of all cases a bookmark or a publication
was copied from another user.\footnote{We ignore imports of bookmark or publication lists (\eg browser bookmark or
  \BibTeX{} files) because during such transfers of own collections to \bibs,  it would not be
  meaningful to look for resources in other users' collections.}
Users copied publications (\publCopy\%) more often than bookmarks
(\bookCopy\%). One reason for this difference might be the fact, that
users leave the system when they follow a bookmarked link, while they
stay within \bibs when they check out details of a publication. Thus,
using \eg a bookmarklet provided by \bibs is the easiest way to post a
website and clicking the copy button on the page is more likely an
option for a publication.  We note, that the share of \bookCopy\% of
copied bookmarks is close to the $2.2\%$ share reported
by Millen and Feinberg~\cite{Millen2006} for the IBM-internal system \dogear, while the
share for publications (\publCopy\%) exceeds that value by a factor of
eight.

Since a resource could only be copied if another user had already posted
that resource in \bibs, we have to take into account whether posted
resources were already present in the system when a user posted them.
Among those posts that were not created by copying from another
user, only about \couldBeCopied\% had the corresponding resource already in the
system and thus could have been copied.
Of all posts that could have been created through copying at the time
of posting, a share of roughly \indeedCopied\% has indeed been copied. This can
be regarded as a relatively large share, since looking up publications or
websites in \bibs is only one out of many possible ways to find interesting bookmarks and publications on the web or elsewhere.

\emph{Copying Tags.} Finally, we study whether not only resources, but
also tags are copied. To that purpose we counted how often users who
copied a resource used tags from their own vocabulary or tags of the
original post to describe their new post. In 87\% of all copy
requests at least one tag from the own vocabulary was used. In 42\% of
all copies at least one of the original post's tags was adopted.  In
the other copy events, 44\% of the original posts had only special tags
like ``imported'' that are probably not meaningful for the user
copying the post. Similarly to the copying of resources we thus find
that users copy tags in a large number of cases; although in the majority
of cases own tags were used.

\assumptiondiscussion{} We found evidence for both, personal
information management and social interaction. In general the findings
fit well to the result from~\cite{heckner2009personal} that the
motivation for posting websites and publications is not predominantly
social, \eg the low share of visits to groups and friends. However,
while users might not contribute content particularly to share it
(like in social networks), we could yet observe evidence that they do
profit from the availability of other users' content. The shares of
visited posts and copied resources and tags are evidence of social
interaction and demonstrate, that the collaborative aspect of the
tagging system is recognized and used. For webmasters of such
systems our results confirm, that it is reasonable to assist users in
discovering content of others \eg through search functions or through
recommendations.

%%% Local Variables: 
%%% mode: latex
%%% TeX-master: "paper"
%%% End: 

\begin{figure*}[h!t!]
  \centering
  \subfigure[revisit count (publ.)]{
    \includegraphics[width=0.23\linewidth]{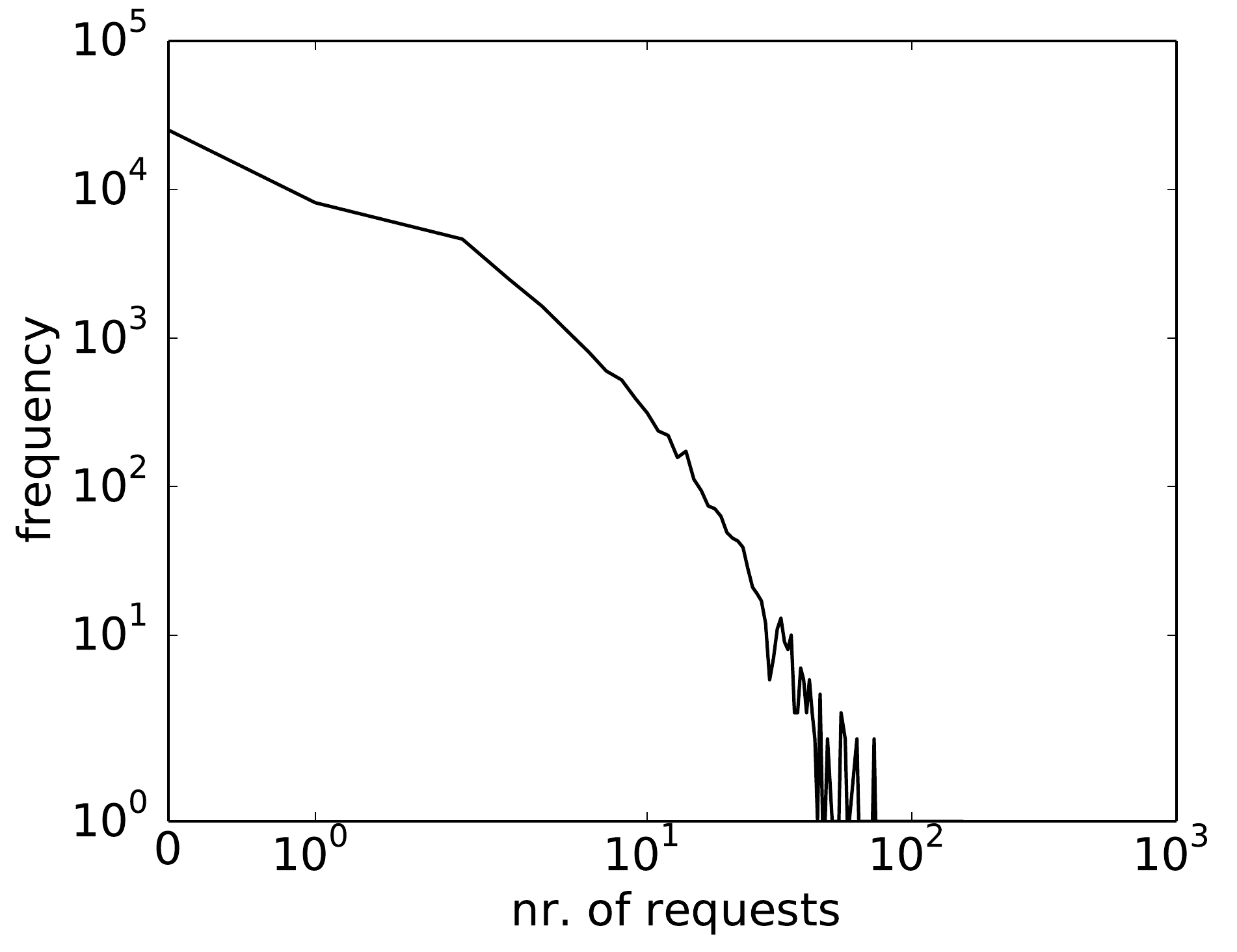}
    \label{fig:revisit_count_res}
  }
  \subfigure[revisit time gap (publ.)]{
    \includegraphics[width=0.23\linewidth]{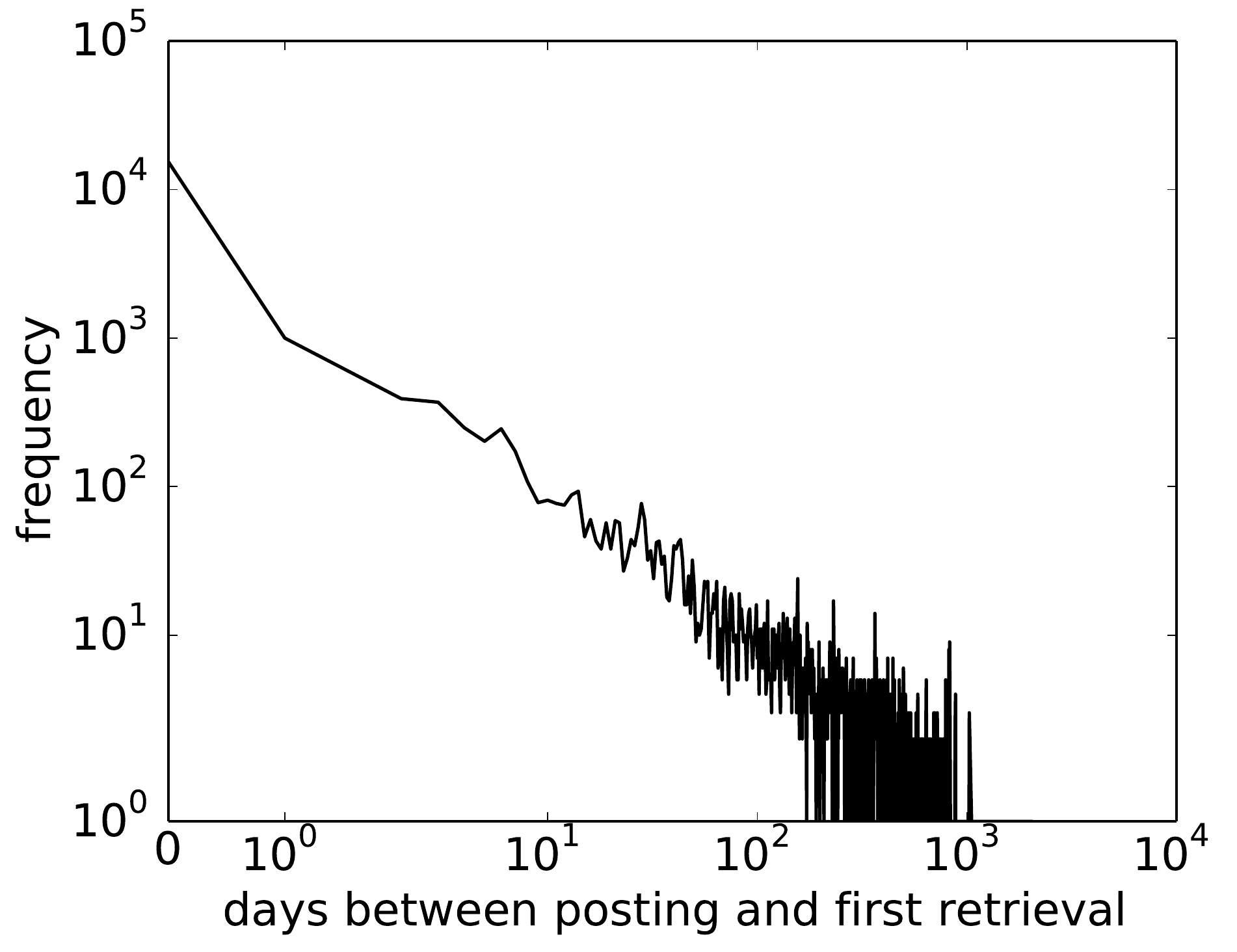}
    \label{fig:revisit_time_res}
  }
  \subfigure[revisit count (tags)]{
    \includegraphics[width=0.23\linewidth]{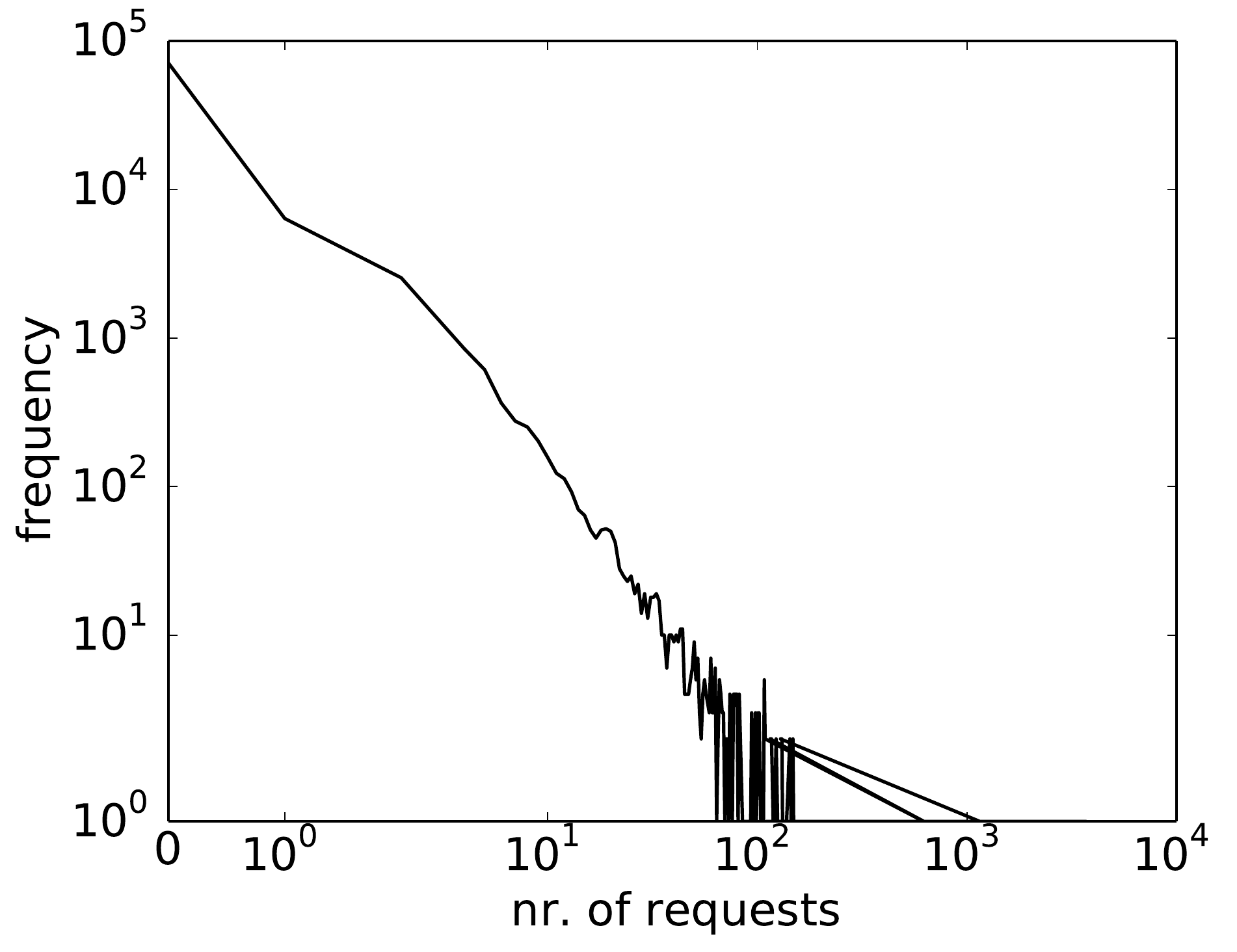}
    \label{fig:revisit_count_tags}
  }
  \subfigure[revisit time gap (tags)]{
    \includegraphics[width=0.23\linewidth]{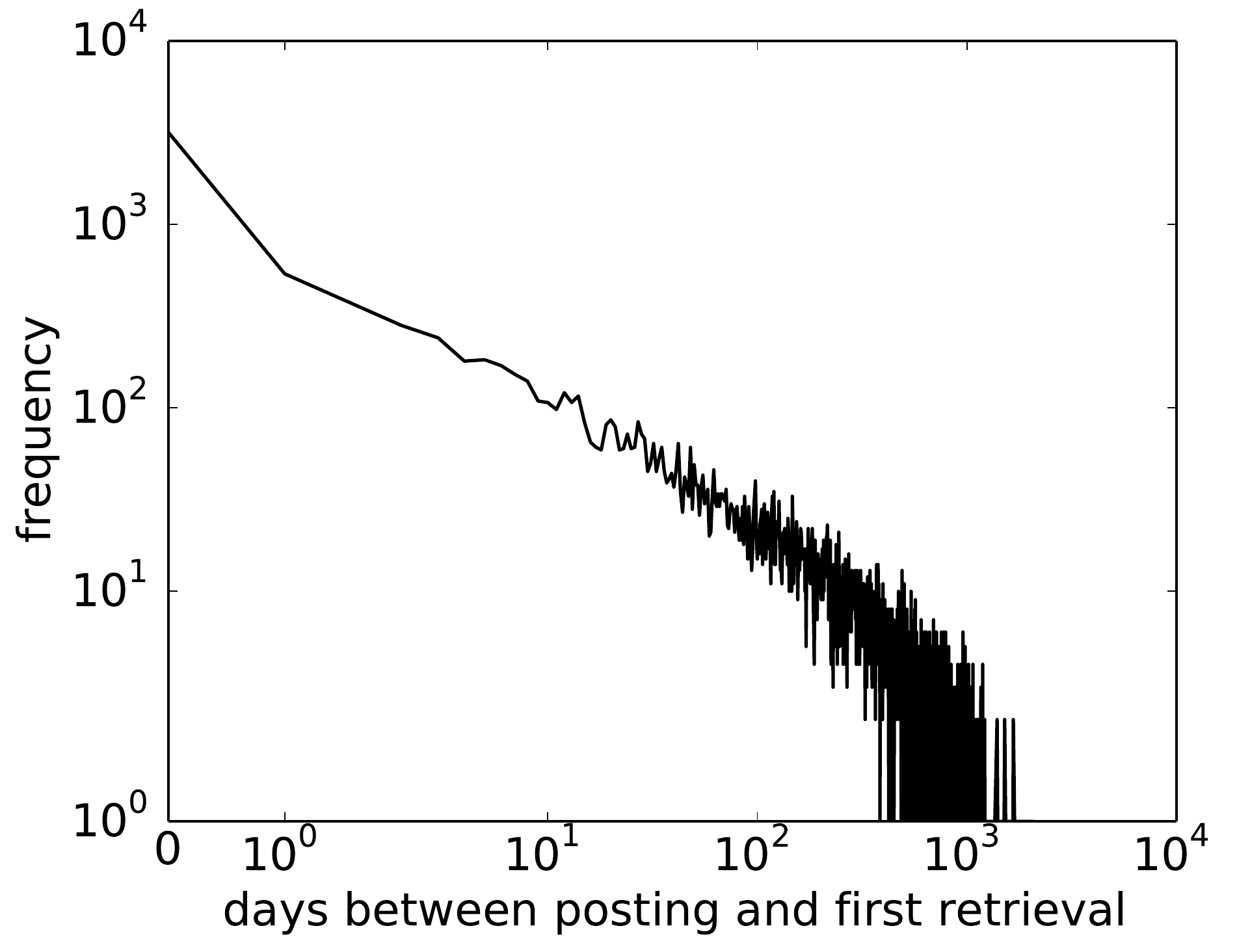}
    \label{fig:revisit_time_tags}
  }
  \caption{\textbf{Revisitation behavior of users.} All four figures are
  visualized on a log-log scale.
  \protect\subref{fig:revisit_count_res}~illustrates the number of times users revisit their
  own publications and \protect\subref{fig:revisit_time_res}~the number of
  days elapsed between the posting of a publication and its first retrieval by its owner.
  \protect\subref{fig:revisit_count_tags}~and~\protect\subref{fig:revisit_time_tags} display the
  revisit count and elapsed days for tags accordingly.
}
   \label{fig:revisit}
 \end{figure*}

\subsection{The Retrieval Assumption}
\assumptionintro{} With the \emph{retrieval assumption}, we refer to the 
notion that tagging systems are used to manage personal collections of resources
for their retrieval later on.

\assumptionevidence{} In a study on the usage of browser bookmarks
by Abrams~\etal{abrams1998information} it was found that users revisit about 96\% of their own
 bookmarks within one year. This gives
rise to the assumption that in tagging systems posts also serve an archival
purpose.
It was hypothesized already at the very beginning of social tagging research
that personal information management may be one of the main reasons why
people use social tagging systems, \eg by
Vander Wal~\cite{VanderWal2005}. Since the idea of social bookmarking can be
seen as an extension of the classic browser based bookmarking and since
we found in the previous section that users spend more requests on their own
collection than on other users' posts, it seems plausible to assume
that users of tagging systems frequently revisit their own posts and tags.
As already mentioned in the foregoing section, the user survey
by Heckner~\etal{heckner2009personal} identified personal management as the main
motivation to post resources to systems like \bibs.

\assumptionresults{} We present statistics about revisiting patterns
obtained for both publication posts\footnote{Requests to bookmarks
  could not be analyzed since they target pages outside \bibs and therefore
  requests for such pages are not recorded in the logs (see
  Section~\ref{sec:dataset}).} and assigned tags. More precisely, we investigate how many times users revisit \emph{their own} posts and tags and also the time difference
between the posting of a resource or tag and its first retrieval, counted in days.
In order to give users a reasonable amount of time for
revisits, we capture all posts until the end of 2010 and all requests
until 2011. This means that each user had at least a whole year
to revisit their posted resources and tags.

The results are shown in Figure~\ref{fig:revisit}. Around 49\% of all
publications were revisited by their owner at least
once.
If a publication has been revisited at all, it mostly was revisited only once
(see Figure~\ref{fig:revisit_count_res}). Furthermore, we can observe in
Figure~\ref{fig:revisit_time_res} that most of the first revisits to a
page took place shortly after the resource had been posted, often on
the same day. These visits could well be control visits to check the
created post, however, it could also mean that users posted a
publication immediately before they used it (\eg in a citation).
The revisit investigations of tags show a more drastic picture. Only
around 17\% of tags are used in queries at least once by a user who
previously assigned them to some post. In
Figures~\ref{fig:revisit_count_tags}~and~\ref{fig:revisit_time_tags} we can observe similar patterns as for
publications. If revisited, tags mostly only have been revisited once and often
shortly after the assignment.

\assumptiondiscussion{} In the previous section we saw that
interactions with the personal collection account for the dominant
share of users' retrieval requests. Although, according to
Heckner~\etal{heckner2009personal} users use the system for later
retrieval, we now find that only about half of all publications are
revisited. Particularly interesting about this observation is that it
does not agree with the work by Abrams~\etal{abrams1998information} on
browser bookmarks, where 96\% of all bookmarked resources got
revisited in the time span of one year. The difference might result from
several factors: First of all, using a publication is different to
revisiting a website -- many websites often renew their content
frequently and are easier to consume than scientific
publications. Moreover, the user survey reported the difficulty of
creating and organizing the bookmarked resources, whereas tagging
systems aim to simplify the process of creating and ordering bookmarks
as much as possible. This could implicate that users tend to store
more, simply because the effort is low. Another reason for the lower
retrieval rate is certainly, that the retrieval of single posts is
only one way to make use of the own collection. Another reasonable way
to use stored publications for citing them is to mass export (\eg
simply all or many publications in the collection) them into a
suitable citation format and selecting the actually used publications
offline. More surprising is the small share of own tags used for
retrieval. An explanation for this observation might be that it is
reasonable to use many tags for a resource to increase the chance of
successful retrieval later on. Furthermore, we will show in the next
section, that using tags is not the dominant way to query the system
anyway.
\Eg for tag recommendation the results indicate, that
the relevance rankings of recommendations should not only take the quantity
of posts into account (\eg like all algorithms in~\cite{jaeschke2008tag}) but
also the visits to them.

%%% Local Variables: 
%%% mode: latex
%%% TeX-master: "paper"
%%% End: 

\subsection{The Equality Assumption}\label{sec:equality}

\story{In the folksonomy model, users, resources, and tags are simply
  classes of entities. TAS are symmetric. The folksonomy tri-graph or
  the folkrank-graph model the tas as links between the nodes. We can
  check wether these links and the entity classes themselves are used
  with equal probabilities or if there are differences. To that end we
  analyze \\ - the absolute requests to entity pages \\ - the relative
  request counts to entity pages (relative to number of entities) \\ -
  transition probabilities in the transition graph.}

\assumptionintro{} The \emph{equality assumption}
\hypertext{states} that the three entity sets in a tagging system --
the sets of users, resources, and tags -- are equally important, \eg for
navigation or retrieval.

\assumptionevidence{} A folksonomy -- the structure underlying tagging
systems -- has been defined as the sets of users $U$, resources
$R$, and tags $T$ together with the tag-assignment-relation $Y
\subseteq U \times R \times T$ (compare~\cite{hotho2006information}). 
In that model, users, resources, and
tags are treated equally and in fact even symmetrically. The
folksonomy model has been widely accepted and many algorithms build on
it, \eg the FolkRank by Hotho~\etal{hotho2006information} or the tensor
factorization method by Rendle~\etal{rendle2009learning}. Since tag
assignments link entities of all three sets together, the idea of the typical
folksonomy navigation is that these entities can be navigated
following these links (\eg clicking a tag to request all posts to
which that tag is assigned to).

\begin{table}[b!]
\centering
\caption{\textbf{Entity request shares in BibSonomy.} We report for each set of folksonomic entities
the average number of requests per entity in that set -- \eg dividing the
total number of requests to tags by the total number of tags -- as well as the total
number and relative share of requests to entities of that set -- among all
requests (total) and among requests targeting content outside the own collection (to others), \eg a request to the user Y by user X.}
%%% This file has been created using the class class org.bibsonomy.analysis.tag.CategoryCounting
\begin{tabular}{l|r|r|r}
 & user & resource & tag \\
\hline
\npnoround
\# requests (per entity) & \nprounddigits{2}\numprint{37.96648449698862} & \nprounddigits{2}\numprint{0.14243862959384282} & \nprounddigits{2}\numprint{1.0887658581791455}\\
\hline
\npnoround
\# requests (total) & \npnoround\numprint{680815} & \npnoround\numprint{327703} & \npnoround\numprint{272566}\\
\nprounddigits{2}\% requests (total) & \nprounddigits{2}\numprint{53.14366583299768} & \nprounddigits{2}\numprint{25.58013369927343} & \nprounddigits{2}\numprint{21.27620046772889}\\
% \hline
% \npnoround
% \# requests (to self) & \npnoround\numprint{567141} & \npnoround\numprint{226903} & \npnoround\numprint{195715}\\
% \nprounddigits{2}\% requests (to self) & \nprounddigits{2}\numprint{57.30091870849368} & \nprounddigits{2}\numprint{22.92507570024622} & \nprounddigits{2}\numprint{19.774005591260096}\\
\hline
\npnoround
\# requests (to other) & \npnoround\numprint{113674} & \npnoround\numprint{100800} & \npnoround\numprint{76851}\\
\nprounddigits{2}\% requests (to other) & \nprounddigits{2}\numprint{39.01965159186476} & \nprounddigits{2}\numprint{34.60053205183215} & \nprounddigits{2}\numprint{26.3798163563031}\\
\end{tabular}
%%% Local Variables:
%%% mode: latex
%%% TeX-master: "../../paper"
%%% End:

\label{tab:categoryCounts}
\end{table}

\assumptionresults{} \emph{Request shares.} Like in the previous
assumptions we analyze retrieval requests. We split them into requests
querying specifically for users, tags, or resources.\footnote{Note,
  that requests to resources are generally underrepresented due to the
  lack of recorded requests to bookmarks (see
  Section~\ref{sec:dataset}).} Hereby, queries with more than one
queried entity have been assigned to the set of that entity that
dominates the request.  For example, the post's details page belongs
to the post's owner, but the target is clearly the resource rather
than the user. A request containing a user and a tag has been counted
as a tag request. Requests that are
not specific to some entity (like the landing page) have been ignored.
\begin{figure}[t!]
 \centering
 \includegraphics[width=0.45\textwidth]{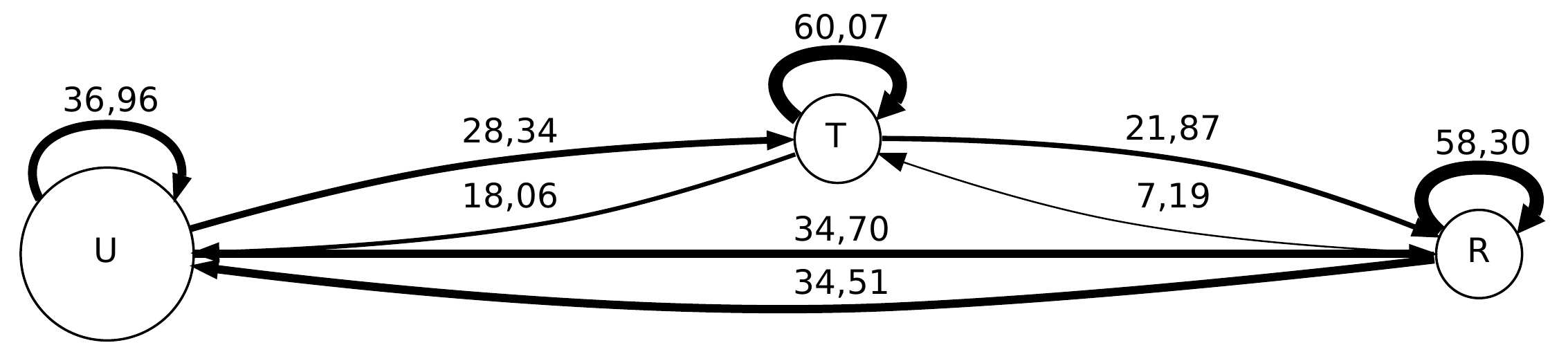}
 \caption{\textbf{Transition probabilities between the three entities
     of the folksonomy.} The nodes \textbf{U}ser,
   \textbf{R}esource and \textbf{T}ag correspond to the columns in
   Table~\ref{tab:categoryCounts} and their size reflect the total
   number of requests to entities of these sets. The edges represent transition
   probabilities from a page of one entity set to another.
     }
 \label{fig:transitiongraph}
\end{figure}

For each set of entities (users, resources, and tags),
Table~\ref{tab:categoryCounts} shows the average number of requests 
to an entity of that set (in the first row) -- 
\eg the number of requests to any tag divided by the total number of tags --
  and the total number of
requests to entities of that set, together with their relative shares (in the
second and third row) compared to the total number of requests. For comparison,
we also report the requests to entities and their shares by only
looking at requests where users have accessed content of other users.
We can clearly see that the number of requests in total
are not equally distributed. There are about \nprounddigits{1}\numprint{2,497798698}
times more requests to specific users than to specific tags, the share of
resources is slightly higher than that of tags. From the average requests per
entity we can deduce that this strong imbalance is not caused simply by a
similar imbalance in the size of the sets. Despite the fact that \bibs has
far more tags and resources than it has users, a user page is queried much more
often than a resource or a tag page on average.

As the use of a tagging system consists of both work with one's own
as well as with posts from other users, we analyze
the same request shares for the latter case separately in rows four
and five of Table~\ref{tab:categoryCounts}.
The share of user requests drops, because we excluded requests to the own user pages
that are accountable for the larger share of requests in \bibs (see the section
on the social assumption). Nevertheless, the queries
for users still outnumber those for tags, however to a lesser extent.
It is also interesting to note that the ratio between requests 
to tags and resources is approximately the same: \numprint{1,20228862} (total requests) vs. \numprint{1,311628996}
(to others). This indicates a comparable user behavior within one's own
collection and within the content of other users.

With above mentioned assignment of each request to one dominating entity, we chose a
  rather conservative approach that tends to underestimate the
  relevance of requests to users. In a similar experiment we directly counted
  the requested entities. Thus a request to a post with requested resource and
  requested user was counted for both user and resource. The result (omitted
  here due to space limitations) shows an even stronger imbalance towards
  users, \ie about \nprounddigits{0}\numprint{65,43705620669004}\% of the
  requested entities were users.

\emph{Transition Probabilities.}
Next, we look at navigational transition probabilities (determined from the
requests' referer
attribute using first order Markov chain probabilities) from
one entity set to another (see Figure~\ref{fig:transitiongraph}). We can observe that
self-transitions are dominant, suggesting that
users tend to stay with the same type of entity in their navigational
paths through \bibs. Aside from that, there are a lot
of transitions from user pages to resource pages and tag pages. This is
not surprising, as user pages consist of listings of a user's
resources, which can be reached with a single click. This also
explains the transitions back to user pages and symbolizes the
``browsing'' in the system. The exception to that is that there are few 
transitions from a resource page to tag pages, 
meaning that users seem only rarely interested in resources with the same tags as the resource at hand.

\assumptiondiscussion{}\label{sec:equality:discussion} We can observe a strong \emph{inequality}
between the use of the three folksonomy entities of users, tags and
resources. While the numbers of requests to tags and to individual
resources are similar, they are dominated by the requests to user
pages. This is surprising, as there are fewer user pages than tag or
resource pages available in \bibs. When discussing navigation within
folksonomies, resources are usually regarded as targets of queries. As
navigational means to find or retrieve theses resources, often tags --
compared to users -- receive the larger interest, as they can
function as resource descriptors. In \bibs, it seems however, that the
user pages are the main means of navigation. From the transition
probabilities, we can find that especially navigation from resources to
tags (and thus to potential further resources to the same tag/topic)
is rather rare. This observation is again surprising. It means, tag
based navigation is less prominent, and algorithms like FolkRank, that
model the transitions between entities, need to be revisited. In
FolkRank, transitions between users, tags, and resources are modeled with equal
probabilities, which -- as we found out -- does not reflect actual user
behavior properly.

%%% Local Variables:
%%% mode: latex
%%% TeX-master: "paper"
%%% End:

\threefig{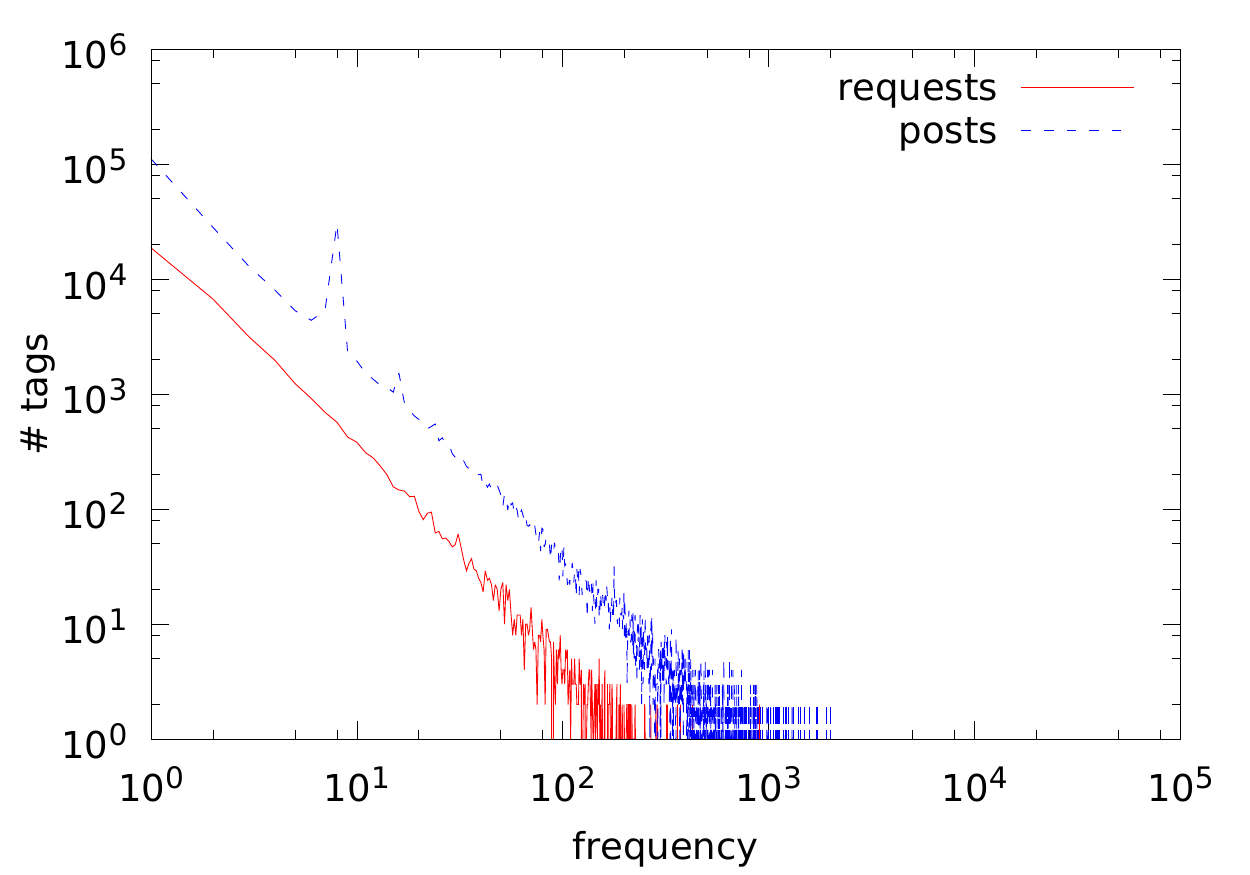}{Distributions
  \distPTagReqFreqAll and \distPTagDBFreqAll}
{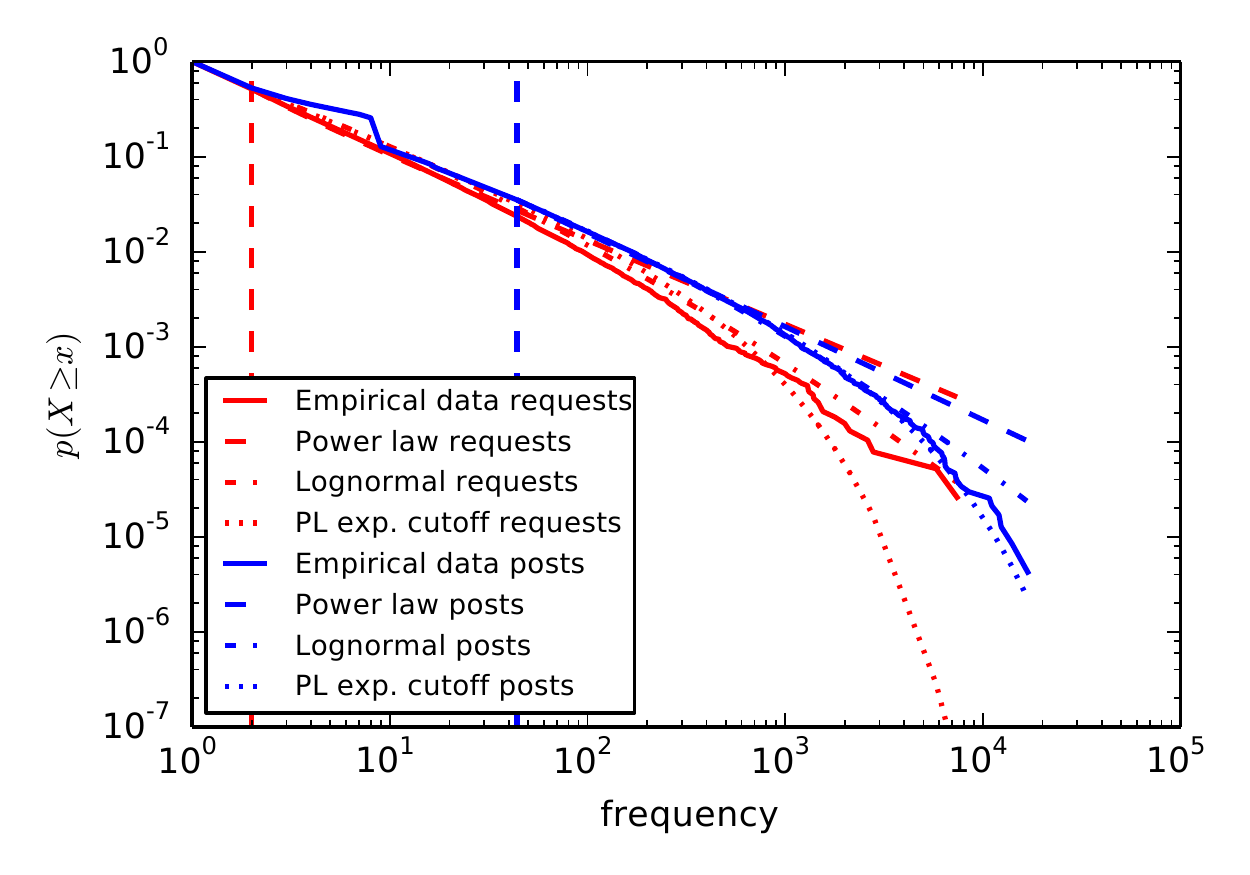}{Different fits to
  \distPTagReqFreqAll and \distPTagDBFreqAll}
{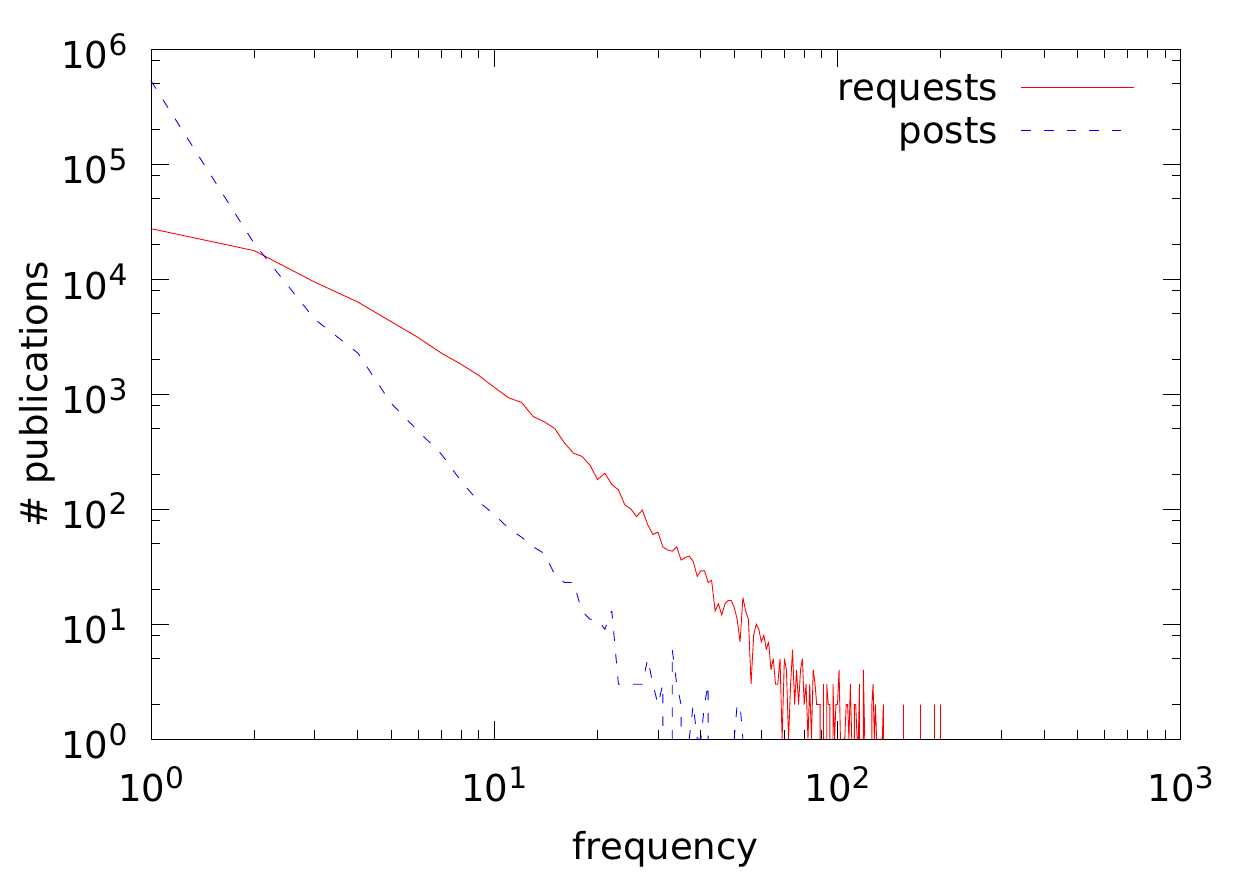}{Distributions
  \distPPubReqFreqAll and \distPPubDBFreqAll}
{logLoggedInNoSpamDBNoSpamTag}{\captioncaption{Frequency distributions
    in requests and posts} 
    In log-log scale, displayed are
    \protect\subref{fig:logLoggedInNoSpamDBNoSpamTagSimple}~the frequency
    distributions \distPTagReqFreqAll and \distPTagDBFreqAll for tags,
    \protect\subref{fig:logLoggedInNoSpamDBNoSpamTagFit}~fits of the respective
    complementary cumulative probability distributions to different standard cumulative probability
    distributions (the vertical lines indicate the corresponding \xmin values), and
    \protect\subref{fig:logLoggedInNoSpamDBNoSpamResBibtexSimple}~frequency
    distributions \distPPubReqFreqAll and \distPPubDBFreqAll for resources.}

\subsection{The Popularity Assumption}
\label{ssec:as:tagcloud}

\assumptionintro{} The \emph{popularity assumption} captures the idea
that the popularity of folksonomic entities -- the number of posts a
user, a resource, or a tag occurs in or its frequency distributions -- matches
similar properties in requests. 

\assumptionevidence{}
In tagging systems, the notion of popularity is exploited in several
ways: (i)~special ``popular'' pages summarize the most frequently
posted resources or tags, (ii)~next to a resource, the number of posts
it occurs in is shown, (iii)~users' profile pages often show the
number of their posts, and (iv)~several algorithms for the
recommendation of tags~\cite{jaeschke2008tag} and
resources~\cite{bogers2009recommender} suggest the most frequently
used entities. Perhaps most prominently, tag frequency is exploited in tag
clouds where the frequency of a tag corresponds to its
font size and particularly rare tags sometimes are not displayed at
all.
Brooks and Montanez~\cite{Brooks2006} point out that it is taken for granted
that the tags a user assigns are the same as those a reader would
select.  Hence, the authors identified the relationship between the
task of article tagging and information retrieval as an open question
to investigate. In the user study by Sinclair and
Cardew-Hall~\cite{sinclair2008folksonomy} it was found that tag clouds are
perceived as visual summaries of resources and that clicking in tag clouds
requires less cognitive effort than entering search queries. This indicates
that the size of a tag is indeed relevant for the users in their query
behavior, but to the best of our knowledge, the correlation between tag usage
in posts and requests has not yet been investigated in a large-scale scenario
other than for the company internal system \dogear~\cite{Millen2006} for which
a correlation of $0.67$ between the frequencies of a tag in posts and in
requests is reported. Regarding the overall behavior,
Cattuto~\etal{cattuto2007network} found that frequencies of entities in posts
follow a heavy-tailed distribution -- mostly clean power law fits.
Power law functions are known to exhibit scale-invariance and are mostly
explained by the Yule process which is also known as preferential attachment.

\assumptionresults{} \emph{Tags.} Since tag clouds are one of the most
popular applications of popularity, we begin the investigation with
tags and their distributions of frequencies in the request logs
(\distPTagReqFreqAll) and in the posts
(\distPTagDBFreqAll).\footnote{Hereby, we ignore posts from two users
who are known to only automatically create posts from publication
catalogues to provide more content in the system.} More
precisely, \begin{itemize}
\item $\distPTagReqFreqAll(k)$ counts how
  many tags have been \emph{requested} exactly $k$ times (\eg $n=\distPTagReqFreqAll(k)$ means, that exactly $n$ tags have been requested exactly $k$ times) and 
\item $\distPTagDBFreqAll(k)$
  counts how many tags have been \emph{assigned} to exactly $k$ posts (and thus
  constitutes the usual node degree distribution described
  in~\cite{cattuto2007network}).
\end{itemize}
Both distributions are shown in
Figure~\ref{fig:logLoggedInNoSpamDBNoSpamTagSimple}.\footnote{ 
A close investigation of the notable peak in the distribution
\distPTagDBFreqAll{} at frequency \npnoround\numprint{8} reveals, that this
anomaly is due to the activities of one single user, who used
\npnoround\numprint{28989} tags exactly \npnoround\numprint{8} times. We
therefore ignore the peak in the following discussion.}
\nprounddigits{3}
\begin{table}[b!]
\centering
\caption{\captioncaption{Correlation and Divergence of request and tag distributions} Pearson's correlation coefficient $\pears$, Spearman's rank
  correlation coefficient \spear and the Jensen-Shannon divergence \js{2} for
  pairs of distributions. In each row, a distribution \distPEntityReqEntity{}
  (Entity is either tag, user, or resource) of requests (or their
  frequencies (\distPEntityReqFreq)) is compared to a distribution
  \distPEntityDBEntity{} of posts (or their frequencies
  (\distPEntityDBFreq)).}
%%% This table was created using class org.bibsonomy.analysis.tag.PrintCorrellationTable
\begin{tabular}{c|c||c|c|c}
requests & posts & \pears & \spear & \js{2}\\
%\hline
\hline
\distPTagReqFreqAll & \distPTagDBFreqAll & \logLoggedInNoSpamDBNoSpamTagFreqAllPearsonCorrelation & \logLoggedInNoSpamDBNoSpamTagFreqAllSpearmanCorrelation & \logLoggedInNoSpamDBNoSpamTagFreqAllJS \\
%\hline
\distPTagReqEntityAll & \distPTagDBEntityAll & \logLoggedInNoSpamDBNoSpamTagEntityAllPearsonCorrelation & \logLoggedInNoSpamDBNoSpamTagEntityAllSpearmanCorrelation & \logLoggedInNoSpamDBNoSpamTagEntityAllJS \\
%\hline
\distPTagReqEntityNonZero & \distPTagDBEntityNonZero & \logLoggedInNoSpamDBNoSpamTagEntityNonZeroPearsonCorrelation & \logLoggedInNoSpamDBNoSpamTagEntityNonZeroSpearmanCorrelation & \logLoggedInNoSpamDBNoSpamTagEntityNonZeroJS \\
%\hline
\hline
\distPUserReqFreqAll & \distPUserDBFreqAll & \logLoggedInNoSpamDBNoSpamUserFreqAllPearsonCorrelation & \logLoggedInNoSpamDBNoSpamUserFreqAllSpearmanCorrelation & \logLoggedInNoSpamDBNoSpamUserFreqAllJS \\
%\hline
\distPUserReqEntityAll & \distPUserDBEntityAll & \logLoggedInNoSpamDBNoSpamUserEntityAllPearsonCorrelation & \logLoggedInNoSpamDBNoSpamUserEntityAllSpearmanCorrelation & \logLoggedInNoSpamDBNoSpamUserEntityAllJS \\
%\hline
\distPUserReqEntityNonZero & \distPUserDBEntityNonZero & \logLoggedInNoSpamDBNoSpamUserEntityNonZeroPearsonCorrelation & \logLoggedInNoSpamDBNoSpamUserEntityNonZeroSpearmanCorrelation & \logLoggedInNoSpamDBNoSpamUserEntityNonZeroJS \\
%\hline
\hline
\distPPubReqFreqAll & \distPPubDBFreqAll & \logLoggedInNoSpamDBNoSpamResBibtexFreqAllPearsonCorrelation & \logLoggedInNoSpamDBNoSpamResBibtexFreqAllSpearmanCorrelation & \logLoggedInNoSpamDBNoSpamResBibtexFreqAllJS \\
%\hline
\distPPubReqEntityAll & \distPPubDBEntityAll & \logLoggedInNoSpamDBNoSpamResBibtexEntityAllPearsonCorrelation & \logLoggedInNoSpamDBNoSpamResBibtexEntityAllSpearmanCorrelation & \logLoggedInNoSpamDBNoSpamResBibtexEntityAllJS \\
%\hline
\distPPubReqEntityNonZero & \distPPubDBEntityNonZero & \logLoggedInNoSpamDBNoSpamResBibtexEntityNonZeroPearsonCorrelation & \logLoggedInNoSpamDBNoSpamResBibtexEntityNonZeroSpearmanCorrelation & \logLoggedInNoSpamDBNoSpamResBibtexEntityNonZeroJS \\
\end{tabular}
%%% Local Variables:
%%% mode: latex
%%% TeX-master: "../../paper"
%%% End:

\label{tab:correlation}
\end{table}
The first observation is, that \distPTagDBFreqAll{} dominates
\distPTagReqFreqAll{}, meaning that in total there are more tag
assignments than requests for tags. 
Since tag frequency distributions in posts (\distPTagDBFreqAll{}) are
known to be heavy-tailed~\cite{cattuto2007network} -- mostly power law -- it was
to be expected that the distribution of tag frequencies in the request logs
(\distPTagReqFreqAll{}) has similar properties.
To confirm this, we first fitted the power law function ($y
= cx^{-\alpha}$ where $x > \xmin$) to the empirical data
using the methods of Clauset~\etal{Clauset2009}. Next, we compared the corresponding fit to
the exponential function as a lower barrier for heavy-tailed distributions as
well as other heavy-tailed probability distributions, namely the
lognormal function and the power law function with an
exponential cutoff (which means that for large $x$ values the function
deviates from the typical power law function).
We visualize the empirical distributions, the best power law $\xmin$ values
(vertical lines) and the corresponding fits in
Figure~\ref{fig:logLoggedInNoSpamDBNoSpamTagFit} for both \distPTagDBFreqAll{}
as well as \distPTagReqFreqAll.\footnote{For better visibility we omitted the
(weak) exponential fit.} For the fits of the power law function we obtained
$\alpha=1.98$ and $\xmin=44$ for \distPTagDBFreqAll{}, and $\alpha=1.88$ and
$\xmin=2$ for \distPTagReqFreqAll.
The distributions are similar with regard to their slopes $\alpha$.
Noteworthy is the higher result of $\xmin$ for \distPTagDBFreqAll{} (in
contrast to the small value for \distPTagReqFreqAll{}), indicating that
the power law fit only holds for a smaller portion of the distribution
(the tail). Visual inspection suggests that there are slightly fewer tags with low frequencies than
one would expect in a power law distribution. While an in-depth
analysis of this phenomenon is beyond the scope of this work, we can
speculate that it might be a consequence of the use of tag
recommenders that typically suggest tags that are already frequently
used, leading to an ignorance of low frequency tags.

A comparison between the fits to the different distributions
showed that the power law function is a statistically significantly
better fit to the data than the exponential fit. Both the
lognormal as well as the power law function with an exponential cutoff
are also good fits to the data confirming our assumption about
heavy-tailed distributions and they are even slightly better fits to
the data compared to the pure power law function as one can see in
Figure~\ref{fig:logLoggedInNoSpamDBNoSpamTagFit}.  This can be
explained by the slight decay in the distributions -- visible where
the line of the empirical distribution (\distPTagReqFreqAll{} at
$\approx 10^2$ and \distPTagDBFreqAll at $\approx 10^3$) falls below
the straight line of the respective power law fit. Similar to the
explanations by Mossa~\etal{mossa2002truncation} -- which are also
discussed by Cha~\etal{cha2009analyzing} -- this may be reasoned due
to information filtering which might hinder preferential
attachment. However, we need to keep in mind that there is only a
slight decay visible. Nevertheless, detailed investigations regarding
this cutoff are necessary for a better understanding of this
behavior. 
By and large, we can observe similar processes of how users post tags
and how they request them -- i.e., processes yielding heavy-tailed
distributions.

Further, we directly compare both \distPTagDBFreqAll{} and
\distPTagReqFreqAll{} with each other using Pearson's correlation
coefficient \pears and Spearman's \spear.\footnote{Note, that all correlation
results in this section are statistically significant with a p-value below
$0.05$, which is why we do not directly report it explicitly for each
calculation.} From the first row in Table~\ref{tab:correlation} we can observe
that the Pearson's and Spearman's correlations are high. An explanation for the
smaller Spearman's \spear value is the fluctuation in the distributions (see
Figure~\ref{fig:logLoggedInNoSpamDBNoSpamTagSimple}) where the number of tags
no longer decreases monotonously with increasing frequency. Finally, a
comparison of the distributions using the Jensen-Shannon divergence \js{2}
confirms similarity.

In the tag frequency distributions, we found similarity in the way \emph{how}
users use and request tags. As a next step, we analyze the tag
popularity on the level of individual tags, to see whether there are
similarities regarding \emph{which} tags users assign and
request. Particularly, we look at the distributions
\distPTagReqEntityAll and \distPTagDBEntityAll, where
\begin{itemize}
\item $\distPTagReqEntityAll(t)$ is the number
of \emph{requests} to a tag $t$ (\eg  $n=\distPTagReqEntityAll(\text{"web"})$ means, the tag ``web'' has been requested exactly $n$ times) and 
\item $\distPTagDBEntityAll(t)$ is the number of \emph{posts} that the tag $t$
occurs in.
\end{itemize}

\begin{figure}[t!h!]
  \centering

\includegraphics[width=0.35\textwidth]{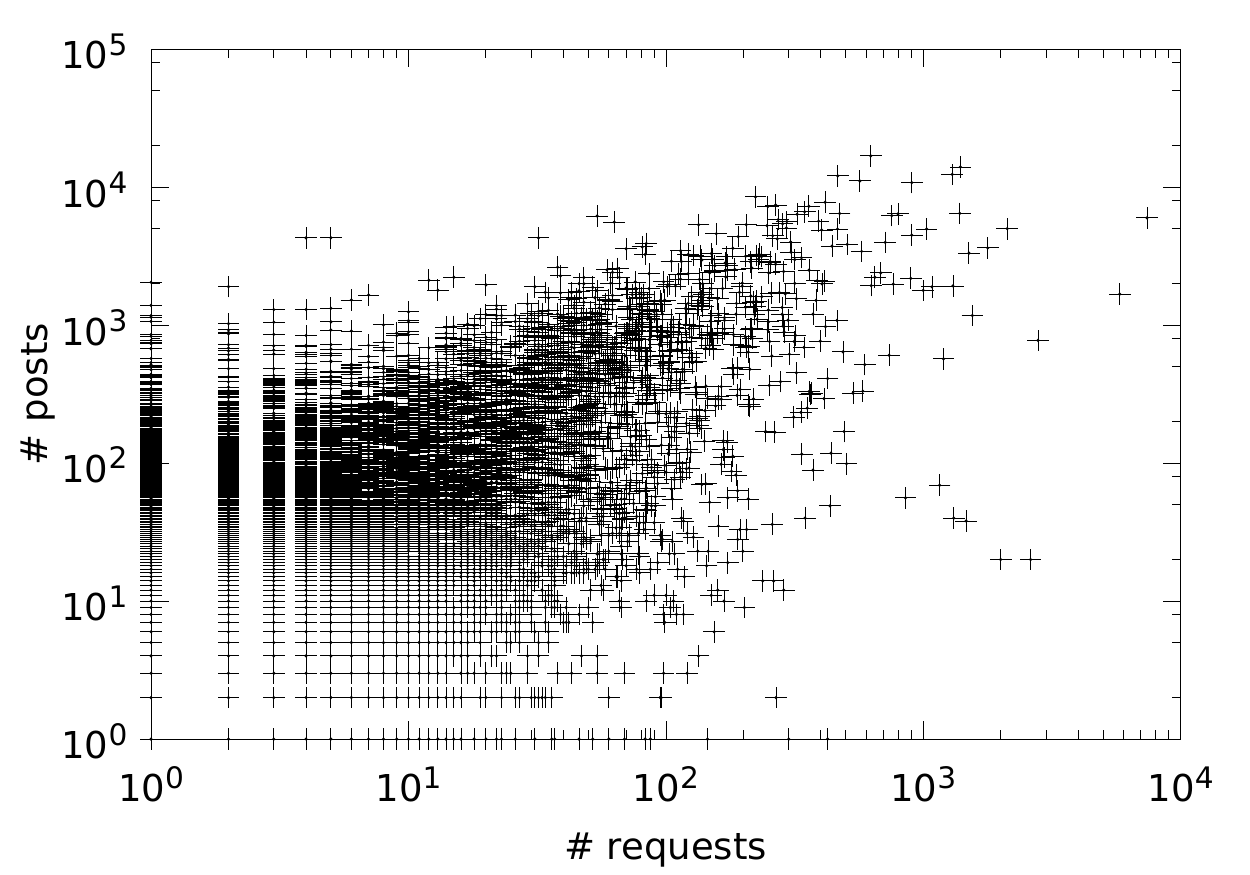}
\caption{\captioncaption{Occurrences of tags in requests and in posts} The scatter plot in log-log scale of the numbers of requests to a tag
 $t$ vs. the number of post a tag $t$ occurs in. Only for higher frequencies, the number of requests and posts of a tag appear to be correlated.
}
  \label{fig:logLoggedInNoSpamDBNoSpamTagEntityScatter}
\end{figure}
Figure~\ref{fig:logLoggedInNoSpamDBNoSpamTagEntityScatter} shows the
scatterplot of these two tag distributions where each point
in the diagram denotes one tag $t$ with its number of requests
$\distPTagReqEntityAll(t)$ and its number of posts
$\distPTagDBEntityAll(t)$ as coordinates. We can immediately see that
despite the similarity in the behavior of tag frequencies, there are enormous
differences on the level of individual tags.
Only for very frequent tags (more than 100 requests) one could presume
a correlation between both frequency counts. To quantify the effect,
the second row of Table~\ref{tab:correlation} shows the correlation
coefficients and the Jensen-Shannon divergence for the two
distributions $\distPTagReqEntityAll(t)$ and
$\distPTagDBEntityAll(t)$. Different than for the previous
distributions we can observe rather low correlation and a much higher
divergence. This means -- contrary to the popularity assumption --
that the number of posts a tag is assigned to and the number of times
a tag is queried are only mildly correlated. The found correlation of
$\pears=0.42$
is also lower than the one reported for the company system Dogear
($0.67$).

A closer look at the log data revealed, that many tags which have been
used in posts were never queried at all and several tags have been
queried but were never assigned to any post. Therefore, we look at
similar distributions as before but we specifically ignore tags that
only occur in one of the two tag distributions.
We yield distributions \distPTagReqEntityNonZero{} and
\distPTagDBEntityNonZero, reducing the number of considered tags
significantly to only \nprounddigits{0}$\numprint{10,55266255}\%$. 
Their distributions' correlations and divergence can be found in the
third row of Table~\ref{tab:correlation}. We can observe that the
limitation to such ``active'' tags yields higher Spearman correlation
and less divergence,
as the active tags' rankings exhibit far less ties than the full set of
tags.

\emph{Users and Publications.} As with tags, we investigated similar
distributions of both users and resources, \ie \distPUserReqEntityAll{}
counting the requests to specific users, \distPUserDBEntityAll{} counting a
user's posts, \distPPubReqEntityAll{} counting the requests to a particular
publication and \distPPubDBEntityAll{} counting the posts containing a
publication. Similarly, we have the according frequency distributions (\eg
\distPUserReqFreqAll) and the restricted distributions to active entities
ignoring those that occur either only in posts or only in requests (\eg
\distPPubReqEntityNonZero). Hereby again, we restrict resources to publications
(and thus omit bookmarks), as visits of bookmarks are not recorded in the log
files (see Section~\ref{sec:dataset}).
The correlation results are depicted in rows four through nine in
Table~\ref{tab:correlation} and for publications the frequency distributions
 are illustrated in
Figure~\ref{fig:logLoggedInNoSpamDBNoSpamResBibtexSimple} (further figures have been omitted due to space
limitations).

The distributions of user (publication) frequencies in requests
\distPUserReqFreqAll{} (\distPPubReqFreqAll{}) and in posts
\distPUserDBFreqAll{} (\distPPubDBFreqAll{}) are similar and yield
relatively high correlation according to Pearson's~\pears{}
(Table~\ref{tab:correlation}, rows four and seven). Their
Jensen-Shannon divergences~\js{2} are higher than for tags, but still
the distributions are relatively similar. Since the distributions
\distPPubDBFreqAll{} and \distPPubReqFreqAll{} are for the most part
monotonically decreasing
(Figure~\ref{fig:logLoggedInNoSpamDBNoSpamResBibtexSimple}), their
rank correlation is high (unlike for the frequencies of users).
Notable in both cases (users and publications) is that
the distributions of frequencies in posts and requests are no longer
``parallel'' as they were in the case of tags (compare
Figures~\ref{fig:logLoggedInNoSpamDBNoSpamTagSimple}
and~\ref{fig:logLoggedInNoSpamDBNoSpamResBibtexSimple}). 

Power law fits for the publication frequency 
distributions of both posts \distPPubDBFreqAll{} ($\alpha=3.15$,
$\xmin=6$) and requests \distPPubReqFreqAll{} ($\alpha=3.00$, $\xmin=14$)
are decent fits with relatively low $\xmin$ values. Not surprisingly, the fits of
the power law function are statistically significantly better than those of the
exponential function. However, it is extremely difficult to
distinguish the fits of the lognormal function and the power law function with
exponential cutoff from the power law fit -- a strong indicator for the presence
of heavy-tailed distributions.
For user frequencies, our results also indicate a good power law fit
for both \distPUserDBFreqAll{} ($\alpha=2.39$, $\xmin=988$) and
\distPUserReqFreqAll{} ($\alpha=1.58$, $\xmin=3$). Similar to our
investigations on tag frequencies, we obtain a higher $\xmin$ value
\hypertext{for the frequencies in posts than for those in requests. For \distPUserReqFreqAll{} all candidate functions are better fits than the exponential function; both the lognormal as well as the power law function with exponential cutoff are better fits to the data than the pure power law function. The power law with cutoff is even better than the lognormal. For \distPUserDBFreqAll{} the power law fit is better than the exponential function and it is difficult to distinguish from the other}
candidate distributions.

Regarding individual entities, we again measure correlations between
the respective distributions in Table~\ref{tab:correlation} (for users
in rows five and six and for publications in rows eight and nine): For
the resources (\distPPubReqEntityAll{} and \distPPubDBEntityAll{}), we
obtain similar results as previously for tags: Pearson's correlation
is moderate, the divergence is even higher than for tags, there is
almost no rank correlation, and removing ``inactive'' publications
(occurring either only in posts or in requests) yields higher rank
correlation and lower divergence. The elimination of such publications
leaves only about \nprounddigits{0}\numprint{12.19226691}\% 
of the original set of publications. By and large, we find only moderate
correlation even among the actively posted and requested publications. 
A possible explanation for the correlation results might be based on the large
number of publications that only get posted and requested infrequently. Slight
changes in the post or request counts (\eg once vs. twice) only change Pearson's
correlation slightly, but have a large influence on Spearman's correlation.
For users (\distPUserReqEntityAll{} and \distPUserDBEntityAll{}) we
find different behavior: almost no correlation according to Pearson's
\pears, moderate rank correlation \spear (higher than for tags and
publications) and divergence \js{2}. This indicates that
users with many posts indeed tend to be requested more, but not
proportionally more.

\assumptiondiscussion{} The obtained results do not clearly support the
initial assumption. The overall behavior of tag (and to a smaller
degree of user and resource) frequencies is similar in requests and
posts and they are heavy-tailed as expected. In all examples we can
find a good power law fit. However, in some occasions the distribution
decays from the straight power law function which indicates the
presence of other heavy-tailed distributions which might be based on
distinct processes creating these distributions. This warrants further
detailed investigations in the future.

On the level of individual entities, we observe weaker correlations and
only among the more actively used entities. It is surprising that
despite the fact that tag clouds are displayed in \bibs and users can
click tags to find according resources, the choice of tags in
requests is not stronger correlated to their popularity in posts. 
Also, we noted a strong difference to the company internal system
\dogear{} where much stronger correlations could be observed for tags.
For operators of a tagging system, the results indicate
that it is reasonable to exclude rarely requested tags completely from tag
clouds or to use request frequencies instead or in addition to post frequencies
in tag clouds. These could even be personalized to a user's query behavior.

%%% Local Variables: 
%%% mode: latex
%%% TeX-master: "paper"
%%% End: 

%%% Local Variables: 
%%% mode: latex
%%% TeX-master: "paper"
%%% End: 

\section{Discussion}
\label{sec:discussion}
In the previous section we have shown evidence indicating
  to what degree the four discussed assumptions do or do not hold in
  \bibs. While our findings in this paper are limited to \bibs, our
  approach is directly applicable to other tagging systems and we briefly
  discuss some aspects of such a transfer here.   Like shown in the user study
  by Heckner~\etal{heckner2009personal} different tagging systems yield
  different characteristics (in their case regarding the users' tagging
  motivation). We can thus assume that similarly, the four discussed
  assumptions in this paper will hold to different degrees in other tagging
  systems and we can speculate about possible influences.

  We have already mentioned the influence of the \emph{degree of
    openness}. In contrast to public, openly available systems,
  company-internal systems can impose certain requirements on their
  users, like the use of real names instead of pseudo\-nyms or
  boundaries for the tags and resources in the system. For example, the
  knowledge whose resources one browses could be a strong influence for
  the social behavior of sharing and visiting. Indeed we have found
  similarity but also pronounced differences between the usage
  behavior in BibSonomy compared to that in \dogear~\cite{Millen2006}
  in our investigation of the social assumption in
  Section~\ref{ssec:socialness} and also in the popularity assumption
  in Section~\ref{ssec:as:tagcloud}.

  Another influence is surely the \emph{type of resources} that are
  bookmarked.  Heckner~\etal{heckner2009personal} have shown, that
  motivations for tagging (sharing or personal information management)
  were different in the systems Youtube (resources are videos) and
  Flickr (images) compared to Delicious (web links) and Connotea
  (publication references). A major difference between those two pairs
  of systems is that the resources like links and publication
  references are taken from other available sources whereas images and
  videos are often published in the respective system for the first
  time. We thus expect that with regard to the social
  assumption we would find similar results on Delicious and Connotea,
  because \bibs allows users to tag web links (like Delicious) and references to
  publications (like Connotea). On the other hand, we can speculate that systems like Youtube
  and Flickr would show different results.
  The \emph{age} of the system is another aspect. All
  three previous log file analyses~\cite{Millen2006,Damianos2007,Millen2007social} report results
  from periods of eight, ten, and twelve month respectively, shortly
  after the systems' creation in 2005.  In contrast to that, our log
   dataset covers a period of six years.
  Finally, the \emph{navigation concept} and the \emph{graphical user
    interface} can play a role. \bibs offers the typical folksonomy
  navigation by always presenting users, resources, and tags as linked
  entities. However, different tagging systems may make different
  design choices, \eg regarding the visibility and accessibility of
  individual entities.

  To investigate these questions further, one would have to conduct
  experiments on logs of other systems as well. However, the bottleneck hereby is
  the availability of such datasets. Therefore, our study is a first step towards analyzing user behavior using log files. We encourage other researchers and
  webmasters of tagging systems to conduct similar studies, using the
  here presented methods, on their tagging systems and to compare
  their results to ours.

%%% Local Variables: 
%%% mode: latex
%%% TeX-master: "paper"
%%% End: 

\section{Conclusions}
\label{sec:conclusions}

In this work we have tested and challenged a number of prominent assumptions
about social tagging systems using a web server log dataset from the system \bibs
 containing posting and requesting data. 
We have thus supplemented previous work -- that has tapped into surveys
and post data to tackle these issues -- by also reflecting actual user behavior
leveraging request data. Our findings paint a rather mixed picture
about the four assumptions studied in this paper: While we find
evidence both for and against the \emph{social assumption}, we also
find that the \emph{retrieval assumption} might not hold for systems
such as \bibs. Our results suggest that the \emph{equality assumption} is wrong
for \bibs, and the \emph{popularity assumption} only holds partially.

Overall, our work contributes (i) a stepping stone for further studies
on social tagging systems that require request log data and (ii) a
basis for comparative studies, \eg exploring the extent to which
these different assumptions hold in different tagging
systems. It is reasonable to assume that different tagging systems
(such as Flickr, Delicious, \bibs and others) exhibit unique
characteristics and dynamics that make them amenable to different uses
and purposes. Further studies of request log data in other tagging
systems would be helpful in uncovering these differences. In addition, we
provide (iii) new insights about the relative importance of users,
tags and resources in social tagging systems. Finding that the
equality assumption does not hold generally has important implications
for the layout of tagging systems and for the design and
implementation of algorithms that address search and retrieval. For
example, the FolkRank~\cite{hotho2006information} algorithm might
profit from the inclusion of weights reflecting popularity or
transition probability in requests.

We hope our work triggers a new line of research on social tagging
systems that utilizes traces of actual user behavior, to test and
challenge our existing body of knowledge about these systems gained
from other inquisition methods, such as surveys or post data.

%%% Local Variables:
%%% mode: latex
%%% TeX-master: "paper"
%%% End:

\section{Acknowledgments}
 This work is in part funded by the FWF Austrian Science Fund Grant
 I677 as well as by the DFG through the PoSTS project.

\bibliographystyle{abbrv}
\balance
\bibliography{bibliography}
\end{document}